\documentclass[longauth]{aa} 
\usepackage[colorlinks=true,linkcolor=blue,citecolor=blue]{hyperref}
\usepackage[dvipsnames]{xcolor}
\usepackage{graphicx}
\graphicspath{{graphics}}
\usepackage{txfonts}
\usepackage[separate-uncertainty=true, list-units=repeat]{siunitx}
\usepackage{subcaption}
\usepackage[inline]{enumitem}
\usepackage{multirow}
\usepackage{bm}

\newcommand{\vandokkumlabel}{JWST-ER1}

\newcommand{\COSMOSWeb}{COSMOS-Web}
\newcommand{\COSMOS}{COSMOS}

\newcommand{\JADES}{JADES}
\newcommand{\ALPINEALMA}{ALPINE-ALMA [C\textsc{ii}]}

\newcommand{\sextractor}{\textsc{SourceXtractor++}}
\newcommand{\SX}{\textsc{SE++}}
\newcommand{\Lephare}{\textsc{LePhare}}
\newcommand{\Cigale}{\textsc{Cigale}}
\newcommand{\EAZY}{\textsc{EAZY}}
\newcommand{\slfit}{\textsc{sl\_fit}}
\newcommand{\pyautolens}{\textsc{PyAutoLens}}

\newcommand{\HST}{\textit{HST}}
\newcommand{\JWST}{\textit{JWST}}
\newcommand{\HSC}{HSC}
\newcommand{\UVISTA}{UltraVISTA}
\newcommand{\SCUBA}{SCUBA-2}
\newcommand{\Herschel}{\textit{Herschel}}

\newcommand{\DES}{DES}
\newcommand{\Subaru}{\textit{Subaru}}
\newcommand{\AGEL}{AGEL}
\newcommand{\ALMA}{ALMA}
\newcommand{\NIRCAM}{NIRCam}
\newcommand{\MIRI}{MIRI}
\newcommand{\SPIRE}{SPIRE}
\newcommand{\PACS}{PACS}

\DeclareSIUnit{\Msun}{M_\odot}
\DeclareSIUnit{\year}{yr}
\DeclareSIUnit{\pc}{pc}
\DeclareSIUnit{\mag}{mag}
\DeclareSIUnit{\mas}{mas}
\DeclareSIUnit{\dex}{dex}
\DeclareSIUnit{\jansky}{Jy}
\DeclareSIUnit{\kpc}{kpc}

\newcommand{\Cring}{COSMOS-Web ring}
\newcommand{\ClumpEast}{CE}
\newcommand{\ClumpWest}{CW}

\begin{document} 

   \title{The \COSMOSWeb{} ring: in-depth characterisation of an \\ Einstein ring lensing system at $z \sim 2$}
   \subtitle{}  

   \author{W.~Mercier\inst{\ref{LAM}}\fnmsep\thanks{\email{wilfried.mercier@lam.fr}}
           M.~Shuntov\inst{\ref{DAWN},\ref{NBI}} \and%
           R.~Gavazzi\inst{\ref{LAM},\ref{IAP}} \and%
           J.~W.~Nightingale\inst{\ref{DurhamComputation}} \and%
           R.~Arango\inst{\ref{LAM}} \and%
           O.~Ilbert\inst{\ref{LAM}} \and%
           A.~Amvrosiadis\inst{\ref{DurhamComputation}} \and%
           L.~Ciesla\inst{\ref{LAM}} \and%
           C.~Casey\inst{\ref{UAT}, \ref{DAWN}}\and%
           S.~Jin\inst{\ref{DAWN},\ref{DTU}}\and%
           A.~L.~Faisst\inst{\ref{Caltech}} \and%
           I.~T.~Andika\inst{\ref{TUM}, \ref{MPI}} \and%
           N.~E.~Drakos\inst{\ref{HawaiiHilo}} \and%
           A.~Enia\inst{\ref{Bologna}, \ref{INAF}} \and%
           M.~Franco\inst{\ref{UAT}} \and%
           S.~Gillman\inst{\ref{DAWN}, \ref{DTU}} \and%
           G.~Gozaliasl\inst{\ref{Helsinki}, \ref{Aalto}} \and%
           C.~C.~Hayward\inst{\ref{NewYorkComputation}} \and%
           M.~Huertas-Company\inst{\ref{IAC}, \ref{LERMA}, \ref{Paris-Cite}, \ref{laLaguna}} \and%
           J.~S.~Kartaltepe\inst{\ref{Rochester}} \and%
           A.~M.~Koekemoer\inst{\ref{STScI}} \and%
           C.~Laigle\inst{\ref{IAP}} \and%
           D.~Le~Borgne\inst{\ref{IAP}} \and%
           G.~Magdis\inst{\ref{DAWN}, \ref{DTU}, \ref{NBI}} \and%
           G.~Mahler\inst{\ref{DurhamComputation}, \ref{DurhamExtragal}} \and%
           C.~Maraston\inst{\ref{Portsmouth}} \and%
           C.~L.~Martin\inst{\ref{SantaBarbara}} \and%
           R.~Massey\inst{\ref{DurhamComputation}, \ref{DurhamExtragal}} \and%
           H.~J.~McCracken\inst{\ref{IAP}} \and%
           T.~Moutard\inst{\ref{LAM}} \and%
           L.~Paquereau\inst{\ref{IAP}} \and%
           J.~D.~Rhodes\inst{\ref{NASA}} \and%
           B.~E.~Robertson\inst{\ref{UnivCalifornia}} \and%
           D.~B.~Sanders\inst{\ref{HawaiiHonolulu}}\and%
           M.~Trebitsch\inst{\ref{Groningen}} \and%
           L.~Tresse\inst{\ref{LAM}} \and%
           A.~P.~Vijayan\inst{\ref{DAWN}, \ref{DTU}}
    }

   \institute{Aix Marseille Univ, CNRS, CNES, LAM, Marseille, France\label{LAM}%
    \and%
    Cosmic Dawn Center (DAWN), Denmark\label{DAWN}%
    \and%
    Niels Bohr Institute, University of Copenhagen, Jagtvej 128, 2200 Copenhagen, Denmark \label{NBI}%
    \and%
    Institut d’Astrophysique de Paris, UMR 7095, CNRS, Sorbonne Université, 98 bis boulevard Arago, F-75014 Paris, France\label{IAP}%
    \and%
    Institute for Computational Cosmology, Durham University, South Road, Durham DH1 3LE, UK\label{DurhamComputation}%
    \and%
    DTU Space, Technical University of Denmark, Elektrovej, Building 328, 2800, Kgs. Lyngby, Denmark\label{DTU}%
    \and%
    Caltech/IPAC, 1200 E. California Blvd. Pasadena, CA 91125, USA\label{Caltech}%
    \and%
    The University of Texas at Austin, 2515 Speedway Blvd Stop C1400, Austin, TX 78712, USA\label{UAT}%
    \and%
    Instituto de Astrofísica de Canarias (IAC), La Laguna, E-38205, Spain\label{IAC}%
    \and%
    Observatoire de Paris, LERMA, PSL University, 61 avenue de l’Observatoire, F-75014 Paris, France\label{LERMA}%
    \and%
    Université Paris-Cité, 5 Rue Thomas Mann, 75014 Paris, France\label{Paris-Cite}%
    \and%
    Universidad de La Laguna. Avda. Astrofísico Fco. Sanchez, La Laguna, Tenerife, Spai\label{laLaguna}%
    \and%
    Institute for Astronomy, University of Hawaii, 2680 Woodlawn Drive, Honolulu, HI 96822, USA\label{HawaiiHonolulu}%
    \and%
    Department of Astronomy and Astrophysics, University of California, Santa Cruz, 1156 High Street, Santa Cruz, CA 95064 USA\label{UnivCalifornia}%
    \and%
    Institute of Cosmology and Gravitation, University of Portsmouth, PO13FX, Portsmouth, UK\label{Portsmouth}%
    \and%
    Centre for Extragalactic Astronomy, Durham University, South Road, Durham DH1 3LE, UK\label{DurhamExtragal}%
    \and%
    Department of Physics and Astronomy, University of Hawaii, Hilo, 200 W Kawili St, Hilo, HI 96720, USA\label{HawaiiHilo}%
    \and
    Center for Computational Astrophysics, Flatiron Institute, 162 Fifth Avenue, New York, NY 10010, USA\label{NewYorkComputation}%
    \and%
    Department of Physics, Faculty of Science, University of Helsinki, 00014 Helsinki, Finland\label{Helsinki}%
    \and%
    Department of Computer Science, Aalto University, PO Box 15400, Espoo, FI-00076, Finland\label{Aalto}%
    \and%
    Technical University of Munich, TUM School of Natural Sciences, Department of Physics, James-Franck-Str. 1, D-85748 Garching, Germany\label{TUM}%
    \and%
    Max-Planck-Institut f\"{u}r Astrophysik, Karl-Schwarzschild-Str. 1, D-85748 Garching, Germany\label{MPI}%
    \and%
    University of California, Santa Barbara\label{SantaBarbara}%
    \and%
    Jet Propulsion Laboratory, California Institute of Technology, 4800 Oak Grove Drive, Pasadena, CA 91109\label{NASA}%
    \and%
    Laboratory for Multiwavelength Astrophysics, School of Physics and Astronomy, Rochester Institute of Technology, 84 Lomb Memorial Drive, Rochester, NY 14623, USA\label{Rochester}%
    \and%
    Space Telescope Science Institute, 3700 San Martin Drive, Baltimore, MD 21218, USA\label{STScI}%
    \and%
    Kapteyn Astronomical Institute, University of Groningen, P.O. Box 800, 9700AV Groningen, The Netherlands\label{Groningen}%
    \and%
    University of Bologna - Department of Physics and Astronomy “Augusto Righi” (DIFA), Via Gobetti 93/2, I-40129, Bologna, Italy\label{Bologna}%
    \and%
    INAF - Osservatorio di Astrofisica e Scienza dello Spazio, Via Gobetti 93/3, I-40129, Bologna, Italy\label{INAF}
   }

   \date{Received; accepted}

 \abstract{}{%
 We provide an in-depth analysis of the \Cring{}, an Einstein ring at $z \approx 2$ that we serendipitously discovered during the data reduction of the \COSMOSWeb{} survey and that could be the most distant lens discovered to date.
 }{%
 We extract the visible and near-infrared photometry of the source and the lens from more than 25 bands. We combine these observations with far-infrared detections to study the dusty nature of the source and we derive the photometric redshifts and physical properties of both the lens and the source with three different SED fitting codes. Using \JWST{}/\NIRCAM{} images, we also produce two lens models to (i) recover the total mass of the lens, (ii) derive the magnification of the system, (iii) reconstruct the morphology of the lensed source, and (iv) measure the slope of the total mass density profile of the lens.
 }{%
 We find the lens to be a very massive elliptical galaxy at $z = 2.02 \pm 0.02$ with a total mass within the Einstein radius of $M_{\rm tot}(<\theta_{\rm Ein}) = \SI{3.66(0.36)e11}{\Msun}$ and a total stellar mass of $M_\star = 1.37^{+0.14}_{-0.11} \times \SI{e11}{\Msun}$. We also estimate it to be compact and quiescent with a ${\rm{sSFR}} \simeq  \SI{e-13} - \SI{e-15}{\per\year}$. Compared to stellar-to-halo mass relations (SHMRs) from the literature, we find that the total mass of the lens within the Einstein radius is consistent with the presence of a dark matter (DM) halo of total mass $M_{\rm h} = 1.09^{+1.46}_{-0.57} \times \SI{e13}{\Msun}$.
In addition, the background source is a $M_\star = \SI{1.26(0.17)e10}{\Msun}$ star-forming galaxy (${\rm{SFR}} \approx \SI{78(15)}{\Msun\per\year}$) at $z=5.48 \pm 0.06$. The morphology reconstructed in the source plane shows two clear components with different colors. Dust attenuation values from SED fitting and nearby detections in the FIR also suggest that the background source could be at least partially dust-obscured.}{
 We find the lens at $z \approx 2$. Its total, stellar, and DM halo masses are consistent within the Einstein ring, so we do not need any unexpected changes in our description of the lens such as changing its initial mass function or including a non-negligible gas contribution. 
 The most likely solution for the lensed source is at $z \approx 5.5$. Its reconstructed morphology is complex and highly wavelength dependent, possibly because it is a merger or a main sequence galaxy with a heterogeneous dust distribution. 
 }


   \titlerunning{The \COSMOSWeb{} ring}
   \authorrunning{Mercier et al.}

   \maketitle
%

\section{Introduction}


Lensing happens whenever a light ray from a background galaxy is deviated towards the observer by a foreground mass distribution, including galaxy clusters \citep[e.g.][]{Lynds1986, Soucail1987, Kneib1996, Campusano2001, Jauzac2015, Massey2018, Richard2021, Claeyssens2022, Atek2023} and massive galaxies \citep[e.g.][]{Walsh1979, Jauncey1991, Dye2014, Nightingale2023, Etherington2023}. Depending on the alignment between the lens and the background source, as well as the shape of the lens' potential, strong lensing can either produce multiple images of the same source \citep[e.g. an Einstein cross,][]{Einstein_cross}, gravitational arcs \citep[e.g.][]{Soucail1987}, or a full Einstein ring \citep[e.g][]{Jauncey1991}. Strong lensing is a powerful tool to study the properties of galaxies for two main reasons. First, it magnifies the flux of the background galaxy, allowing the detection of intrinsically fainter and/or higher redshift sources. In addition, this magnification enhances the resolution of the background source, allowing spatially resolved studies to be undertaken on small structures such as star-forming clumps \citep[e.g.][]{Swinbank2007, Swinbank2009, Jones2010, Livermore2012, Livermore2015, Johnson2017b, Johnson2017, Mevstric2022, Claeyssens2023}. Second, the magnification and the shape of the gravitational arcs or rings are primarily impacted by the mass distribution of the lens. Therefore, lensing is one of the few observations along with galaxy dynamics allowing to constrain the total mass content of galaxies, including hidden components such as their dark matter (DM) halo \citep[e.g.][]{Treu2010,Oguri2014, Nightingale2023, Bolamperti2023}. 

For instance, this led to the discovery of the so-called "bulge-halo" conspiracy where the total density profile (i.e. baryons + dark matter) of local early-type galaxies (ETGs) exhibit a nearly isothermal behavior \citep{Gavazzi2007, Auger2010, Etherington2023}. Samples up to $z_{\rm lens} \sim 0.8$ further reveal varying trends in the value of the density profile slope $\gamma$ with increasing $z_{\rm lens}$, from a mild decrease \citep{Sonnenfeld2013b} to a more significant increase \citep{Bolton2012}. Furthermore, cosmological simulations suggest a slight rise in $\gamma$ with redshift \citep{Wang2019, Wang2020}, dependent on feedback mechanisms like active galactic nuclei. However, our current understanding of the evolution of the density profiles of ETGs with redshift faces a limitation due to the scarcity of lenses identified at $z \gtrsim 0.8$.

While initially relatively rare, galaxy-galaxy lensing candidates have now become ubiquitous thanks to large and deep imaging surveys in the optical and near-infrared (NIR) such as in the \Subaru{} Hyper Suprime-Cam (\HSC{}) survey \citep[e.g.][]{wong_survey_2018}, the Hubble Space Telescope (\HST{}) observations of the \COSMOS{} field \citep[e.g.][]{Faure2008b, Faure2008, Jackson2008, pourrahmani_lensflow_2018}, the Dark Energy Survey \citep[\DES{}, e.g.][]{Diehl2017, Rojas2022, odonnell_dark_2022}, the Sloan Digital Sky Survey \citep[SDSS, e.g.][]{Auger2010,Talbot2021, Talbot2022}, the ASTRO 3D Galaxy Evolution with Lenses (\AGEL{}) survey \citep{Tran2022}, the entire \HST{} archives \citep{garvin_hubble_2022}, or in the radio with the Cosmic Lens All-Sky Survey \citep[CLASS, e.g.][]{Myers2003, Browne2003}, the Herschel Multi-tiered Extragalactic Survey \citep[HerMES][]{Wardlow2013}, the Herschel Astrophysical Terahertz Large Area Survey \citep[Herscel-ATLAS, e.g.][]{Negrello2010, Negrello2017}, or using the South Pole Telescope \citep[SPT, e.g.][]{Vieira2010, Hezaveh2011, Hezaveh2013, Vieira2013}. Currently, the largest catalogues contain up to a few hundred excellent galaxy-galaxy lenses, and this effort was made possible thanks to the numerous developments brought to automatic lens detection algorithms \citep[e.g.][]{gavazzi_ringfinder_2014, Petrillo2017, pourrahmani_lensflow_2018, sonnenfeld_survey_2020, Canameras2020, Savary2022, canameras_holismokes_2023}.

Throughout the last decade, serendipitous discoveries and case-by-case studies of lenses have been pushed toward higher redshifts. So far, the highest redshift lenses ever discovered correspond to those of \citet{Canameras17} and \citet{Ciesla20} both at $z \approx 1.5$ and \citet{Wong14} at $z \approx 1.6$. But a new era is beginning for strong lensing with the advent of \JWST{}. Already, \NIRCAM{} and/or NIRISS observations of lensing clusters, such as Abell 2744 \citep[e.g. see][]{Bergamini2023, Bezanson2022} or SMACS0723 \citep[e.g. see][]{Atek2023}, have detected multiple background sources at $z > 9$ \citep[e.g.][]{Bergamini2023, Atek2023}. Complementarily, NIRSpec observations have allowed the spectroscopic confirmation of multiple galaxies at $z \gtrsim 10$ \citep[e.g.][]{Williams2023, Wang2023, Hsiao2023}. Besides \JWST{}, upcoming new telescopes and facilities with dedicated wide surveys will also play a crucial role in greatly improving the number of detected strong lenses. For instance, roughly 17,000 lenses are predicted in the Nancy Grace Roman Space Telescope 2000 square degree survey \citep[see Sect.\,12.2 and Fig.\,12.7 of][]{Weiner2020}, of the order of 120,000 in the Vera C. Rubin Observatory Legacy Survey of Space and Time \citep[LSST, see Sect.6 and beyond of][]{Collett2015}, and between 95,000 \citep[see Sect.4.4 of][]{Holloway2023} and 170,000 \citep{Collett2015} in the Euclid wide survey.

 While such surveys will slightly extend the redshift range of strong lenses, very few are actually predicted to be discovered at $z > 2$ \citep[e.g. see Fig.6 of][]{Collett2015}. Recently, \citet{Holloway2023} released an in-depth estimation of the detectability of strong lenses in the near-infrared (NIR), including \JWST{} surveys such as the JWST Advanced Deep Extragalactic Survey (\JADES{}) and \COSMOSWeb{}. From purely source-lens alignment considerations, lenses could theoretically be detected up to $z \sim 4$ \citep{Holloway2023}\footnote{Taking a detection probability threshold above 10\%.}. However, when accounting for the magnification and shear of the source, the spatial resolution of \JWST{}, and the limiting depth of the surveys, the predicted maximum redshift for lens detection is not predicted to exceed $z \sim 2$, except in the case of extraordinary cases that will be more likely to happen in ultra-deep pencil-beam surveys such as JADES \citep[see Fig.\,5a of][]{Holloway2023}. 

In this paper, we report the serendipitous discovery of the potentially highest redshift lens ever detected at $z \sim 2$. The system was observed during data reduction of the \COSMOSWeb{} survey in April 2023 and was unambiguously identified as an Einstein ring thanks to the high-resolution optical rest-frame images provided by \JWST{} in multiple bands. While preparing this paper, \citet{vandokkum2023massive} published an independent analysis of this system (dubbed named \vandokkumlabel{}). In this paper, we provide an in-depth analysis of the lens and the source by 
\begin{enumerate*}[label=(\roman*)]
    \item combining \HST{} and \JWST{} data with ground-based observations to constrain precisely the photometric redshift and physical properties of the lens and the source,
    \item performing state-of-the-art mass modeling to constrain precisely the total mass of the lens and study its DM content,
    \item reconstruct the morphology of the lensed source and
    \item studying the dusty nature of the source using complementary FIR-to-radio detections.
\end{enumerate*}

We begin by presenting in Sect.\,\ref{sec:Observations} the observations of the system and the photometry used for the analysis. In Sect.\,\ref{sec:Redshifts}, we discuss the photometric redshift and physical properties of both the source and the lens and in Sect.\,\ref{sec:Lens model} we describe the two methods used to perform the lens modeling. Then, we provide an in-depth complementary analysis of the lens and the source in Sect.\,\ref{sec:Properties} and we conclude in Sect.\,\ref{sec:Conclusions}. 

We adopt a standard $\Lambda$CDM cosmology with $H_0=70$\,km\,s$^{-1}$\,Mpc$^{-1}$, $\Omega_{\rm m}=0.3$, and $\Omega_{\Lambda}=0.7$. Physical parameters are estimated assuming a \citet{Chabrier03} initial mass function (IMF). The magnitudes are expressed in the AB system \citep{Oke74}.

\section{Observations}
\label{sec:Observations}

\subsection{\COSMOSWeb{} ground- and space-based observations}
\label{sec:Observations/COSMOS-Web}

The \Cring{} was discovered thanks to the \JWST{}/\NIRCAM{} imaging of the COSMOS field as part of the COSMOS-Web survey (GO $\#$1727), described in detail in \cite{CWeb2022}. Briefly, it consists of 255-hour imaging of a contiguous 0.54 deg$^{2}$ area in four \NIRCAM{} filters (F115W, F150W, F277W, F444W), down to a 5$\sigma$ depth of AB mag $27.2 - 28.2$, measured in empty apertures of $0.15\arcsec$ radius \citep{CWeb2022}. The \NIRCAM{} data reduction was carried out using the version 1.10.0 of the \JWST{} Calibration Pipeline \citep{Bushouse2022}, Calibration Reference Data System (CRDS) pmap-1075 and a \NIRCAM{} instrument mapping imap-0252. Mosaics are created on both \SI{30}{\mas} and \SI{60}{\mas} pixel scales for all filters. The \NIRCAM{} image processing and mosaic making will be described in detail in Franco et al. (in prep.). COSMOS-Web offers a unique combination of area, depth, and \NIRCAM{}\footnote{The \Cring{} is not part of the \MIRI{} coverage of \COSMOSWeb{}.} resolution.
to carry out searches for distant strong lensing systems. In this paper, we use the currently available data from COSMOS-Web, which is only about $50 \%$ of the total area, imaged over two epochs (January 2023 and April 2023).

We complement the \JWST{} imaging with the wealth of existing ground-based and \HST{}/ACS data, described in detail in Shuntov et al. (in prep.; see also \citealt{Weaver2022} and \citealt{dunlop_ultravista_2016}). We summarize below the dataset used for our work. The $u$ band imaging comes from the CFHT Large Area $U$-band Deep Survey (CLAUDS; \citealt{Sawicki19}), reaching 27.7 mag (5$\sigma$). The ground-based optical imaging is provided by the HSC Subaru Strategic Program (HSC-SSP; \citealt{Aihara18}). We use the HSC-SSP DR3  \citep{Aihara21} in the ultra-deep HSC imaging region in five $g,r,i,z,y$ broad bands and three narrow bands, with a sensitivity ranging 26.5-28.1 mag (5$\sigma$). In addition, we include the reprocessed Subaru Suprime-Cam images with 12 medium bands in optical \citep{Taniguchi07,Taniguchi15}. The \UVISTA{} survey \citep{McCracken12} provides deep NIR imaging in four broad bands $YJHK_s$ and one narrow band (\SI{1.18}{\micro\meter}). Our source falls in one of the ultra-deep stripes of DR5 \citep{dunlop_ultravista_2016}. While being at a lower resolution and less sensitive than \NIRCAM{}, these data are complementary in terms of wavelength coverage. Finally, we also include the \HST{}/ACS F814W band \citep{Koekemoer07} providing a high-resolution image in the $i$-band.

In order to constrain the dust-obscured star-forming activity of the ring (see Sect.\,\ref{sec:Properties/source/dust}), we adopt the deblended FIR photometry in the latest “Super-deblended” catalog of Jin et al. (in prep.). From this catalog, we use measurements at MIPS \SI{24}{\micro\meter}, \Herschel{} \PACS{} \SI{100}{\micro\meter} and \SI{160}{\micro\meter}, \Herschel{} \SPIRE{} \SI{250}{\micro\meter}, \SI{350}{\micro\meter}, and \SI{500}{\micro\meter} (with conservative lower limits of confusion noise used for the \SPIRE{} bands{}; see \citealt{Nguyen2010}), and \SCUBA{} \SI{850}{\micro\meter} which are obtained by performing the "Super-deblending" technique \citep{Jin2018,Liu2018} with priors from the COSMOS2020 catalog \citep{Weaver2022cosmos2020}. We note that the ring is not selected by COSMOS2020 due to its faintness. The only prior adopted for the deblending is on the position of the $z \sim 2$ lens. Given the size of the ring is approximately $\SI{2}{\arcsec}$, it is not resolved in any of the FIR images mentioned above. Therefore, it is appropriate to use the position of the lens as a prior to deblend the FIR emission from other nearby sources.

\begin{figure}
    \centering
    \includegraphics[width=0.95\hsize]{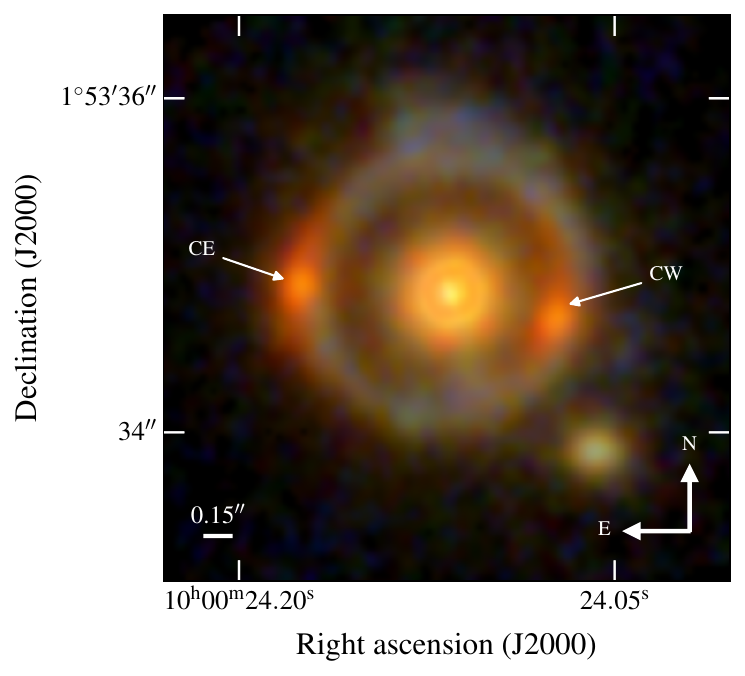}
    
    \caption{False color image of the \Cring{} made by  cutouts from the F814W, F115W, and F150W bands into the blue channel, F277W band into the green channel and F444W band into the red channel. The white bar on the bottom left represents the PSF FWHM in the F444W band. Two clumps \ClumpWest{} and \ClumpEast{} in the ring are identified on the figure. North is up and East is left.}
    \label{fig:Observations/color-image}
\end{figure}

\subsection{Overview of the Einstein ring}
\label{sec:Observations/discovery}

Upon reduction of the \JWST{}/\NIRCAM{} images of \COSMOSWeb{} in April 2023, the \Cring{} was clearly identified by visual inspection. An RGB color image of the system produced by combining PSF-matched \HST{} and \JWST{} cutouts is shown in Fig.\,\ref{fig:Observations/color-image}. The lens is surrounded by the Einstein ring that shows strong inhomogeneities in color. Cutouts of the system in two \JWST{}/\NIRCAM{} bands are shown in the leftmost column of Fig.\,\ref{fig:Observations/morphology} (see also \ref{fig:Appendix/cutouts} for cutouts in all ground- and space-based bands). The morphology of the ring strongly varies with wavelength. This includes \ClumpWest{} (Clump West) and \ClumpEast{} (Clump East) that are detected mostly in F277W and F444W bands, as well as a blue component that is visible throughout the ring, but in particular in its northern and southern parts. The latter looks quite smooth in F277W and F444W bands but it becomes clumpy in F814W, F115W, and F150W bands. Another galaxy is also located roughly \SI{1.2}{\arcsec} South West of the ring and is estimated with \Lephare{} \citep{Arnouts02, Ilbert06} to be at $z = 2.1 \pm 0.1$ . Therefore, this galaxy could potentially be a satellite of the lens, in which case it would be located at around $10 \pm \SI{0.1}{\kilo\pc}$ away from it\footnote{Proper kilo parsec; uncertainty estimated by varying the redshift in the range $1.8 - 2.0$.}. Finally, a last UV-bright galaxy, only visible in $u$, $g$, and $r$ bands is visible West of the ring. The fact that it is not seen near \ClumpEast{} suggests that it is not part of the ring.

Besides \HST{} and \JWST{} images, the system is also detected in all \UVISTA{} bands, as well as in the \HSC{}-$i$, $z$, and $y$ bands (see Fig.\,\ref{fig:Appendix/cutouts}). 
It is interesting to note that the component of the ring that is visible in these bands corresponds to the northern blue part. On the other hand, \ClumpWest{} and \ClumpEast{} do not appear at all. Starting from and blue-ward of the $r$-band, the whole system becomes a dropout. Assuming it corresponds to the Lyman break for the background source, it would place it at $4.5 \lesssim z_{\rm{drop}} \lesssim 6.5$. The lack of detection for the central lens blue-ward of the $r$-band is likely the combination of an intrinsic drop of the spectrum at these wavelengths combined with a coarser spatial resolution\footnote{The median seeing is around \SI{0.9}{\arcsec} in the $u$-band \citep{Sawicki19} and \SI{0.6}{\arcsec} in the $i$-band \citep{Aihara21}} since the CFHT-$u$ band is technically deeper than the HSC-$g$, $r$, and $i$ bands.

When cross-correlating the position of the Einstein ring with strong-lensing catalogues that overlap with the \COSMOS{} field, including that from \citet{Faure2008}, \citet{pourrahmani_lensflow_2018}, \citet{wong_survey_2018}, \citet{sonnenfeld_survey_2018, sonnenfeld_survey_2020}, and \citet{garvin_hubble_2022}, we do not find any source that matches this system. The combination of its dropout nature at low wavelengths and the coarse spatial resolution of the ground-based observations certainly prevented its identification. Still, it is interesting to note that the lens and the ring are clearly visible and can be separated in the \HST{}/F814W band if the image is properly re-scaled to enhance its contrast, and as such, it could have been detected in previous visual searches of strong-lensing candidates in the \COSMOS{} field. Only the \JWST{}/\NIRCAM{} images provide enough signal-to-noise ratio (S/N) to unambiguously identify this system as an Einstein ring. Similarly, \JWST{}/\NIRCAM{} images are mandatory if one wants to derive precise photometric redshifts of both the lens and the source.


\subsection{Photometry}
\label{sec:photometry}

\begin{figure*}
    \centering
    \includegraphics[width=0.95\hsize]{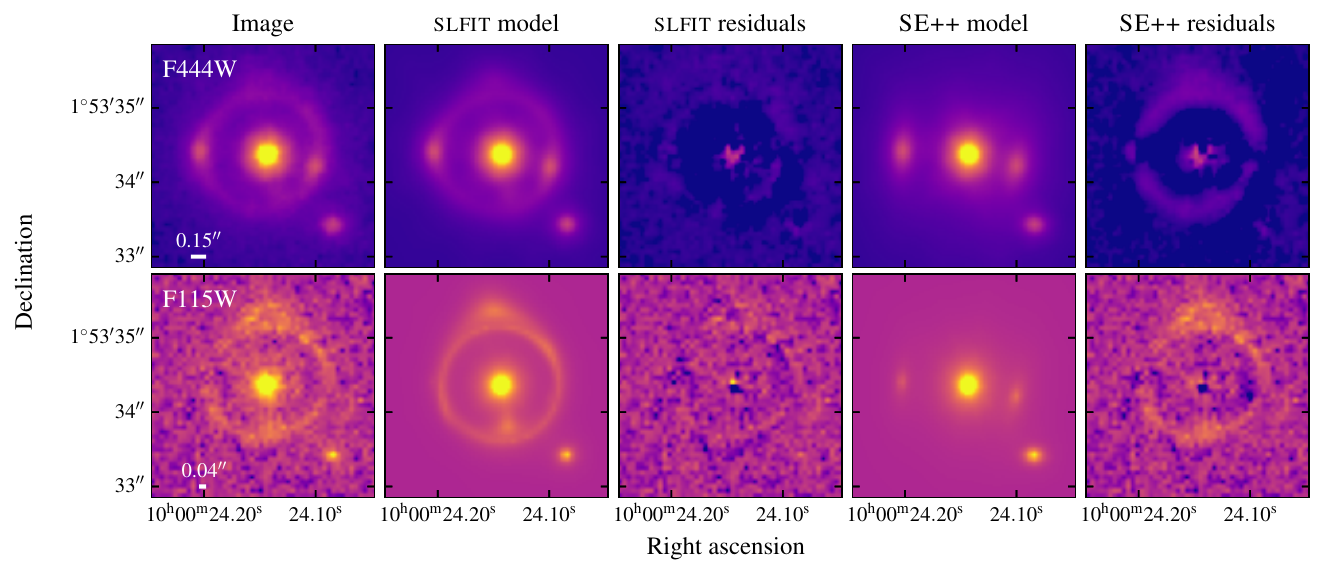}
    
    \caption{Morphology of the \Cring{} in F444W (top row) and F115W (bottom row) bands. From left to right: $\SI{3}{\arcsec} \times \SI{3}{\arcsec}$ cutouts, best-fit lens model in image plane from \slfit{}, residuals from \slfit{}, best-fit morphological model of the lens, \ClumpWest{}, \ClumpEast{}, and the nearby satellite with \sextractor{}, and its residuals. The PSF FWHM is indicated as a white bar on the bottom left corner of the cutout.
    }
    \label{fig:Observations/morphology}
\end{figure*}

We have used two sets of photometry to estimate the redshifts and galaxy properties of this system. The first one (noted \SX{} hereafter) is derived using the \sextractor{} software \citep{bertin20,kummel20}. Four objects in this system were extracted from the full COSMOS-Web catalogue (Shuntov et al., in prep): the lens, the potential satellite, and the two clumps \ClumpWest{} and \ClumpEast{}. We model each galaxy assuming the surface brightness follows a single Sérsic profile. The structural parameters (centre position, Sérsic index $n$, radius, and axis ratio) are constrained using high-resolution \HST{} and \NIRCAM{} images. For other bands (e.g. ground-based), only the flux of each object is allowed to vary during the fit. For each band, the profiles are convolved with the appropriate PSF. Furthermore, the fluxes of the four objects are fitted simultaneously, which is necessary when the objects are blended (e.g. the clumps and the lens in ground-based images).

The second set of photometry is obtained from the lens modeling (hereafter \slfit{} photometry) performed with \slfit{} (see Sect.\,\ref{sec:Lens model/slfit}). The morphology of the lens is modeled with a circular Sérsic profile with a fixed index $n=3$, whereas the source is best modeled with three Sérsic profiles with a fixed index $n = 1$ that do not share the same center. The structural parameters (center position and radius for the lens; center position, radius, and ellipticity for three components of the source) are determined from \JWST{}/\NIRCAM{} images only. Only the flux of each component is allowed to vary when fitting other bands and, as for \SX{} photometry, the appropriate PSF is taken into account for each band during the fit. Examples of \slfit{} and \SX{} photometry extraction for \JWST{}/\NIRCAM{} F444W and F115W bands are shown in Fig.\,\ref{fig:Observations/morphology}. For the source photometry, in our analysis, we use the total flux of all three components as modeled by \slfit{}, unless stated otherwise.

Thus, we always measure the total flux of the lens with \SX{} and \slfit{} photometries. For the source, we measure its total flux with \slfit{} photometry but only the flux in the clumps \ClumpWest{} and \ClumpEast{} with \SX{} photometry. In other terms, \SX{} photometry misses a fraction of the flux of the source found in the remaining parts of the ring. As discussed later, this does not impact the photometric redshifts but it can affect the physical properties of the source (see Sect.\,\ref{sec:Redshifts/source}). Besides, by construction \slfit{} provides intrinsic fluxes (i.e. magnification corrected) whereas \SX{} photometry provides magnified fluxes. In what follows, all physical properties derived with \SX{} photometry are always magnification corrected using the magnification found by \slfit{}.

\section{Photometric redshifts and physical properties}
\label{sec:Redshifts}

We use three SED fitting codes to measure the redshift of the lens and the source. First, we start with the template-fitting code \Lephare{} \citep{Arnouts02, Ilbert06}. We adopt a set of templates extracted from \citet{BC03} assuming 12 different star formation histories (SFH; exponentially declining and delayed), as described in \citet{Ilbert15}. For each SFH, we generate templates at $43$ different ages (from 0.05 to \SI{13.5}{\giga\year}). We assume two attenuation curves \citep{Calzetti00, Arnouts2013} with $E(B-V)$ varying from $0$ to $0.7$. We add the emission line fluxes with a recipe described in \citet{Saito20}, following \citet{Schaerer09}. The normalisation of the emission line fluxes is allowed to vary by a factor of two (using the same ratio for all lines) during the fitting procedure. The absorption of the intergalactic medium (IGM) is implemented following the analytical correction of \citet{Madau95}.
\Lephare{} provides the redshift likelihood distribution for each object, after a marginalization over the galaxy templates and the dust attenuation. We use it as the posterior redshift probability density function (PDF), assuming a flat prior. The physical parameters are derived simultaneously.

Second, we run \EAZY\ \citep{brammer_eazy_2008} to assess the robustness of the photometric redshift. We use the template set that is derived from the Flexible Stellar Population Synthesis models (FSPS) \citep[specifically QSF 12 v3]{Conroy2009, ConroyGunn2010}, with an updated emission line template from \cite{Carnall2023}. \EAZY\ fits a non-negative linear combination of a set of basis templates to the observed flux densities for each galaxy. The latter are corrected for Milky Way extinction internally in the code and the absorption from the IGM is implemented following the prescriptions of \cite{Madau95}. We set a systematic error floor of 0.02 (fraction of the flux), and do not apply any priors or zero-point corrections.

Finally, we also use \Cigale{}\footnote{\url{https://cigale.lam.fr/}} \citep{Boquien19}, a versatile Bayesian-like analysis code modeling the X-rays to radio emission of galaxies, to derive the physical properties of the different objects, as well as assess the robustness of the photometric redshift. \Cigale{} includes multiple modules to model the SFH, stellar, dust, and nebular emission, as well as the active galactic nuclei (AGN) contribution to the SED. The versatility of the code is based on the ability to build and fit the different models in the context of the energy budget balance between the UV-optical emission, from the contribution of young stars, which is absorbed by dust and re-emitted in IR. In this work, we use the following configuration: the \texttt{sfhNlevels} non-parametric module with a bursty continuity prior that was presented and tested in \citet{Ciesla23} \citep[see also][]{Arango23}, the stellar population models (SSPs) of \citet{BruzualCharlot03}, a \citet{Calzetti00} attenuation law as well as the \texttt{skirtor} \citep{Stalevski16} module to model an AGN contribution.

\subsection{Lens}
\label{sec:Redshifts/lens}

\begin{figure*}
    \centering
    \includegraphics[width=0.95\hsize]{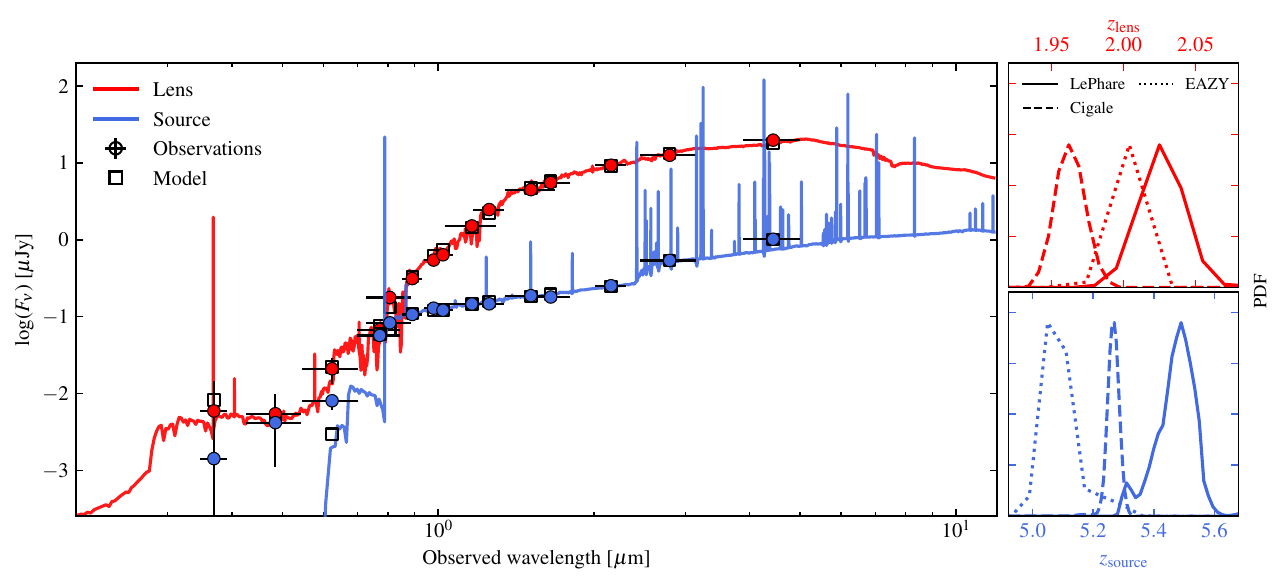}
    
    \caption{\textit{Left:} SED fitting results from \Lephare{} for the lens (red) and the source (blue) using the photometry extracted from the lens modeling with \slfit{} (see Sect.\,\ref{sec:photometry} and \ref{sec:Lens model}). \textit{Right:} redshift PDF for the lens and the source (same colors) when using \Lephare{} (continuous lines), \Cigale{} (dashed lines), and \EAZY{} (dotted lines). We note that the detections in the $u$, $g$, and $r$ bands for the source are likely contaminated by a nearby UV-bright foreground galaxy located West of the ring (see Fig.\,\ref{fig:Appendix/cutouts}.)}
    \label{fig:Redshifts/SEDs}
\end{figure*}

The best-fit \Lephare{} SED model derived on \slfit{} photometry is shown in red on the left panel of Fig.\,\ref{fig:Redshifts/SEDs}. We also show on the right panel of Fig.\,\ref{fig:Redshifts/SEDs} the PDF of the lens' photometric redshift from \Lephare{} (continuous line), \Cigale{} (dashed line), and \EAZY{} (dotted line). The photometric redshift and physical properties derived from the different fits are given in Table\,\ref{tab:Redshift/lens}. We get consistent results between the codes that find the lens at $z \approx 2$. \Lephare{}, \Cigale{}, and \EAZY{} median values agree to within $0.04$ which is also their typical uncertainty. Furthermore, we do not find any significant differences between the values derived from \SX{} and \slfit{} photometries. Given that the morphology of the lens is not affected by the lens model and that \slfit{} and \SX{} photometries are obtained from comparable Sérsic models, it is not surprising that the photometry has little impact. The lens is found to be a massive and quiescent galaxy with $M_\star > \SI{e11}{\Msun}$ and ${\rm{sSFR}} < \SI{e-13}{\per\year}$. More precisely, \Lephare{}, \Cigale{}, and \EAZY{} find respectively $M_{\star} = 1.58^{+0.13}_{-0.12}$, $1.30_{-0.06}^{+0.05}$, and $1.96_{-0.08}^{+0.11} \times \SI{e11}{\Msun}$ with \SX{} photometry and $M_{\star} = 1.37^{+0.14}_{-0.11}$, $1.05^{+0.05}_{-0.05}$, and $1.46^{+0.03}_{-0.03} \times \SI{e11}{\Msun}$ with \slfit{} photometry. We note that the photometric redshift of the lens derived in this study is consistent with the value found in \citet{vandokkum2023massive} and so is the total stellar mass up to a factor of two.

In what follows, we will use the solution from \Lephare{} with \slfit{} photometry at $z_{\rm{lens}} = 2.02$ as a reference and we will discuss how using a different solution might impact our results.

\begin{table}[tbp]
    \centering
    \caption{Photometric redshift estimates and physical properties of the lens.}
    \begin{tabular}{llll}
        \hline
        Code & $z_{\rm{phot}}$ & $M_{\star}$ & SFR \\%
        & & \SI{e11}{\Msun} & \SI{e-2}{\Msun\per\year} \\%
        (1) & (2) & (3) & (4)\\%
        \hline
        \hline
        \Lephare{} (\SX{}) & $1.97^{+0.02}_{-0.02}$ & $1.58^{+0.13}_{-0.12}$ & $<0.01$ \\[2pt]%
        \Cigale{} (\SX{}) & $1.98^{+0.05}_{-0.04}$ & $1.30_{-0.06}^{+0.05}$ & $0.10^{+0.10}_{-0.05}$ \\[2pt]%
        \EAZY{} (\SX{}) & $2.00^{+0.01}_{-0.01}$ & $1.96_{-0.08}^{+0.11}$ & $2.11^{+0.12}_{-0.06}$ \\[2pt]%
        \bf{\Lephare{} (\slfit{})} & \bm{$2.02^{+0.02}_{-0.02}$} & \bm{$1.37^{+0.14}_{-0.11}$} & \bm{$<0.01$} \\[2pt]%
        \Cigale{} (\slfit{}) & $1.96^{+0.08}_{-0.08}$ & $1.05^{+0.05}_{-0.05}$ & $0.38^{+0.65}_{-0.38}$ \\[2pt]%
        \EAZY{} (\slfit{}) & $2.00^{+0.02}_{-0.01}$ & $1.46^{+0.03}_{-0.03}$ & $1.64^{+0.03}_{-0.04}$ \\[2pt]%
        \hline
    \end{tabular}
    \label{tab:Redshift/lens}\\\vspace{5pt}
    {\small\raggedright {\bf Notes:} (1) SED fitting code with, in parentheses, the photometry used, (2) median value, 16th, and 84th quantiles of the photometric redshift PDF, (3) total stellar mass, and (4) star formation rate. Text in boldface represents values used as reference in the analysis.\par}
\end{table}

\subsection{Background source}
\label{sec:Redshifts/source}

The result of the fit with \Lephare{} on \slfit{} photometry is shown in blue on the left panel of Fig.\,\ref{fig:Redshifts/SEDs}. The PDFs from \Lephare{}, \Cigale{}, and \EAZY{} are shown on the right panel and the photometric redshifts and physical parameters for the various fits of the background source are given in Table\,\ref{tab:Redshift/source}. When using \slfit{} photometry, the source is found at $z_{\rm{source}} = 5.48^{+0.06}_{-0.06}$, $5.27^{+0.01}_{-0.03}$, and $5.08^{+0.06}_{-0.04}$ with \Lephare{}, \Cigale{}, and \EAZY{}, respectively. However, with \SX{} photometry, it is found at a lower redshift of $z_{\rm{source}} = 5.27^{+0.02}_{-0.02}$, $4.78^{+0.10}_{-0.15}$, and $5.12^{+0.01}_{-0.01}$ with \Lephare{}, \Cigale{}, and \EAZY{}, respectively. Thus, taking into account uncertainties, the SED fitting codes give us a range of possible photometric redshifts for the background source of $4.63 \lesssim z_{\rm{source}} \lesssim 5.54$. 

We compare our photometric redshift results to that of \citet{vandokkum2023massive}. With our solution at $z_{\rm{source}} = 5.48$ we get a $\chi^2 = 23$. On the other hand, when fitting while fixing the redshift to their solution at $z_{\rm source} = 2.97$, we get a $\chi^2 = 159$. Beside, in the latter case the best-fit SED under-fits in bands around \SI{1}{\micro\meter} and over-fit at both shorter and longer wavelengths. As in \citet{vandokkum2023massive}, if we restrict to \HST{} and \JWST{} bands, both redshifts are valid solutions, though we do get a lower $\chi^2 = 2$ with ours compared to the $\chi^2 = 29$ that we obtain when fixing to their redshift. In other terms, the detection of the source in ground-based data plays a crucial role in the determination of its photometric redshift. If we separate the contributions to the Einstein ring of Comp-1 (i.e. the background component that produces \ClumpWest{} and \ClumpEast{}) and Comp-2 (i.e. the blue component), we get different results. With Comp-1, we find that the solution from \citet{vandokkum2023massive} at $z_{\rm source} = 2.97$ fits slightly better the SED with a $\chi^2$ lowered from 34 (our solution) to 27 (their solution). On the opposite, our solution at $z_{\rm{source}} = 5.48$ is much more robust for Comp-2 with a $\chi^2 = 49$ instead of 900 for their solution. Thus, we believe that the solution of a single background source at $z \approx 5.5$ is consistent, though we cannot completely exclude the possibility that the Einstein ring might actually be the image of two galaxies superimposed along the line-of-sight, one at $z \approx 3$ and another at $z \approx 5.5$. Taking the density of galaxies at these two redshifts in the current version of the \COSMOSWeb{} catalogue, we get a rough estimate on the probability to observe such a superimposition of at most 0.1\%.

All SED fitting solutions find that the source is quite massive with $M_\star \gtrsim \SI{e10}{\Msun}$. For a given SED fitting code, the stellar mass derived from \SX{} photometry is always lower than the value derived from \slfit{} photometry. This is expected since the solution with \SX{} photometry is obtained on \ClumpWest{} only which is just a fraction of the total flux of the ring. With \slfit{} photometry, \Lephare{}, \Cigale{}, and \EAZY{} find a total stellar mass of $M_\star = 1.26^{+0.17}_{-0.16}$, $2.87^{+0.50}_{-0.50}$, and $5.75^{+0.39}_{-0.48} \times \SI{e10}{\Msun}$, respectively. Finally, the SFR of the source is not as precisely constrained as the stellar mass with uncertainties of the order of $5 - \SI{10}{\Msun\per\year}$. Nevertheless, all solutions find that the background source is a star-forming galaxy. Using \slfit{} photometry to estimate the SFR of the entire Einstein ring, \Lephare{} and \Cigale{} find $77.6^{+15.4}_{-11.0}$ and $78.4^{+5.58}_{-5.58}~\unit{\Msun\per\year}$, respectively. Thus, \Lephare{} and \Cigale{} find consistent results with \slfit{} photometry, but also when using \SX{} photometry (i.e. when estimating the SFR of \ClumpWest{} only) with $25.0^{+6.0}_{-3.0}$, and $17.0^{+6.0}_{-5.0}~\unit{\Msun\per\year}$, respectively. Only \EAZY{} finds opposite trends with ${\rm SFR} = 83.0_{-6.00}^{+5.00}~\unit{\Msun\per\year}$ for \SX{} photometry and $24.7_{-2.28}^{+5.52}~\unit{\Msun\per\year}$ for \slfit{} photometry. In other terms, \EAZY{} finds a higher SFR in \ClumpWest{} alone than in the entire Einstein ring. This inconsistency might be the effect of different stellar populations and dust attenuation between the red clumps and the blue part of the ring that \EAZY{} has trouble accounting for (see Sects.\,\ref{sec:Properties/source/multiple galaxies} and \ref{sec:Properties/source/dust}).

In what follows, we will use the solution from \Lephare{} with \slfit{} photometry at $z_{\rm{source}} = 5.48$ as a reference and we will discuss how using a different solution might impact our results.

\begin{table}[tbp]
    \centering
    \caption{Photometric redshift estimates and physical properties of the source. For \SX{}, we quote the results on \ClumpWest{}.}
    \begin{tabular}{llll}
        \hline
        Code & $z_{\rm{phot}}$ & $M_{\star}$ & SFR \\%
        & & $\times \SI{e10}{\Msun}$ & \unit{\Msun\per\year} \\%
        (1) & (2) & (3) & (4)\\%
        \hline
        \hline
        \Lephare{} (\SX{}) & $5.27^{+0.02}_{-0.02}$ & $0.74^{+0.08}_{-0.06}$ & $25.0^{+6.00}_{-3.00}$ \\[2pt]%
        \Cigale{} (\SX{}) &  $4.78^{+0.10}_{-0.15}$ & $1.05^{+0.56}_{-0.66}$ & $17.0^{+6.00}_{-5.00}$ \\[2pt]%
        \EAZY{} (\SX{}) & $5.12^{+0.01}_{-0.01}$ & $1.58_{-0.13}^{+0.11}$ & $83.0_{-6.00}^{+5.00}$  \\[2pt]%
        \bf{\Lephare{} (\slfit{})} & \bm{$5.48^{+0.06}_{-0.06}$} & \bm{$1.26^{+0.17}_{-0.16}$} & \bm{$77.6^{+15.4}_{-11.0}$} \\[2pt]%
        \Cigale{} (\slfit{}) & $5.27^{+0.01}_{-0.03}$ &  $2.87^{+0.50}_{-0.50}$ & $78.4^{+5.58}_{-5.58}$  \\[2pt]%
        \EAZY{} (\slfit{}) & $5.08^{+0.06}_{-0.04}$ & $5.75^{+0.39}_{-0.48}$ & $24.7^{+2.52}_{-2.28}$ \\[2pt]%
        \hline
    \end{tabular}
    \label{tab:Redshift/source}\\\vspace{5pt}
    {\small\raggedright {\bf Notes:} Legend is similar to that of Table\,\ref{tab:Redshift/lens}. Physical parameters from \SX{} photometry correspond to the clump \ClumpWest{} only and are corrected from magnification assuming a magnification factor of $\mu = 11.6$. Values from \slfit{} photometry correspond to the whole ring and are intrinsic by construction. Text in boldface represents values used as reference in the analysis.\par}
\end{table}

\section{Lens modeling}
\label{sec:Lens model}

Lens modeling is an important aspect because
\begin{enumerate*}[label=(\roman*)]
    \item it is the only technique that effectively allows us to recover the intrinsic flux of the whole ring in multiple bands while directly taking into account magnification and distortion from the lens,
    \item it allows us to reconstruct the intrinsic morphology of the background source, and
    \item it gives access to the total mass distribution of the lens.
\end{enumerate*}
In this paper, we have applied two different techniques to model the lens and the deflection of the source. Because they rely on different methodologies and assumptions, we have used them independently and then compared them to assess the reliability of our results. The main difference between these two methods is that the first one fits the source with analytical light profiles (see Sect.\,\ref{sec:Lens model/slfit}), whereas the second one reconstructs the morphology of the source using pixelization (see Sect.\,\ref{sec:Lens model/PyAutoLens}). 

\subsection{Forward light profile fitting with \slfit{}}
\label{sec:Lens model/slfit}

\subsubsection{Method}


\begin{table}[tbp]
    \centering
    \caption{Best-fit parameters for the total mass distribution of the lens using \slfit{} and \pyautolens{} for different redshift solutions of the background source from Table\,\ref{tab:Redshift/source}.}
    \resizebox{0.48\textwidth}{!}{%
    \begin{tabular}{lllll}
        \hline
        Method & $\theta_{\rm{Ein}}$ & $\mu$ & $z_{\rm{source}}$ & $M_{\rm{tot}} (\theta_{\rm{Ein}})$ \\
         & \unit{\arcsec} & & & $\times \SI{e11}{\Msun}$  \\%
        (1) & (2) & (3) & (4) & (5) \\%
        \hline
        \hline
        \multirow{3}{*}{\shortstack{\slfit{}}} & \multirow{3}{*}{\shortstack{$\bm{0.78 \pm 0.04}$}} & \multirow{3}{*}{\shortstack{$\bm{\sim 11.6}$}} & $5.48 \pm 0.06$ & $\bm{3.66 \pm 0.36}$ \\
        & & & $5.27 \pm 0.02$ & $3.75 \pm 0.37$ \\
        & & & $5.08 \pm 0.05$ & $3.84 \pm 0.38$ \\
        \hline
        \multirow{3}{*}{\shortstack{\pyautolens{}}} & \multirow{3}{*}{\shortstack{$0.77 \pm 0.01$}} & \multirow{3}{*}{\shortstack{10.7/14.9}} & $5.48 \pm 0.06$ & $3.56 \pm 0.09$ \\
        & & & $5.27 \pm 0.02$ & $3.64 \pm 0.10$ \\
        & & & $5.08 \pm 0.05$ & $3.73 \pm 0.10$ \\
        \hline
    \end{tabular}}
    \label{tab:sec:Lens model/mass distribution}\\\vspace{5pt}
    {\small\raggedright {\bf Notes:} (1) Method used to model the system, (2) Einstein radius in arc second, (3) average magnification of the source, (4) redshift of the source and its uncertainty used to estimate $M_{\rm{tot}} (\theta_{\rm{Ein}})$, and (5) total mass of the lens within the Einstein radius. The mass is always estimated using $z_{\rm{lens}} = 2.00 \pm 0.02$. Uncertainties on the mass are evaluated by averaging 1000 Monte-Carlo realizations using the uncertainties on the Einstein radius, source redshift, and lens redshift. For \pyautolens{}, we give the two different magnifications found in F444W and F115W, respectively. Text in boldface represents values used as reference in the analysis.\par}
\end{table}

\begin{figure}[htbp]
  \flushright
  \includegraphics[width=1\hsize]{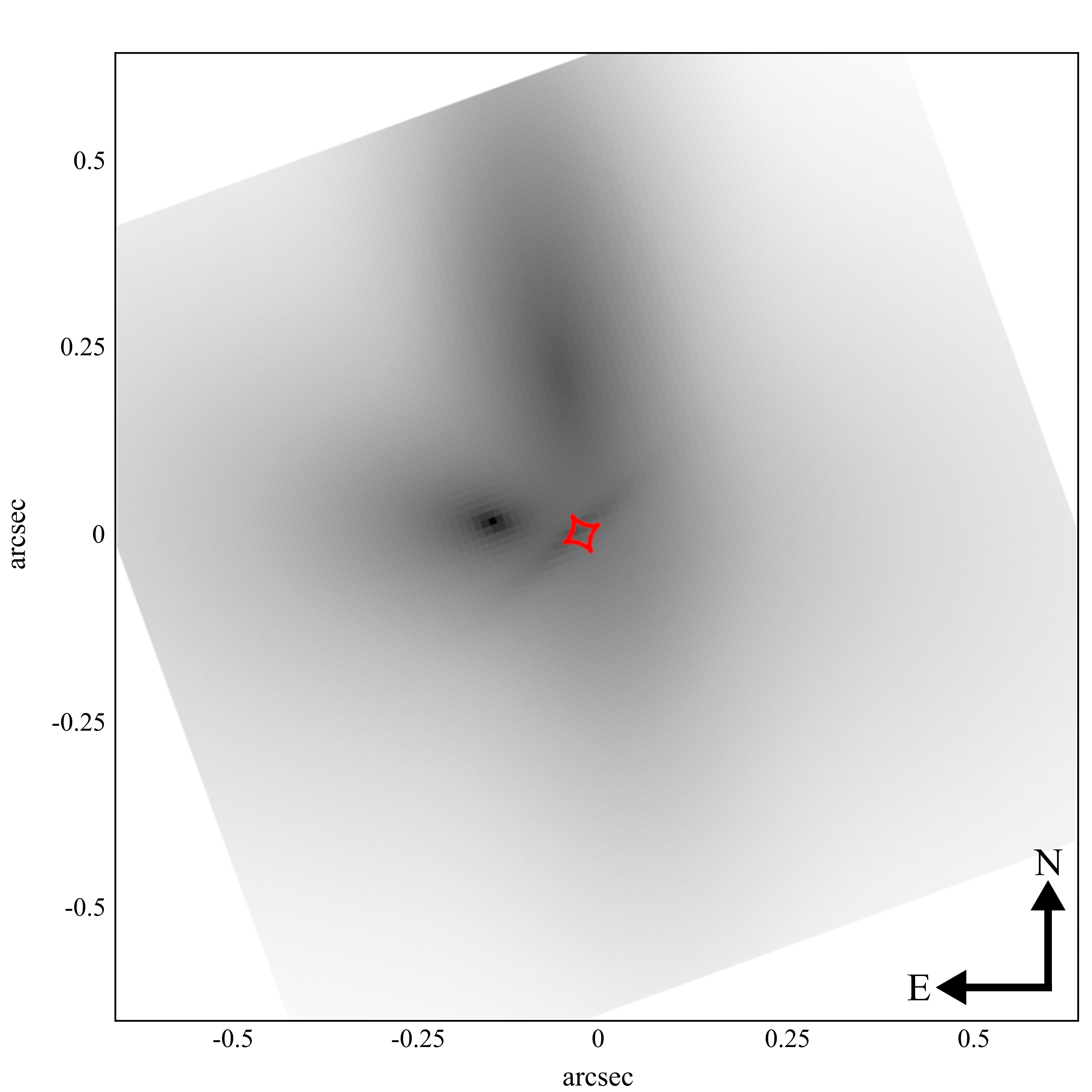}     
  \includegraphics[width=0.9\hsize]{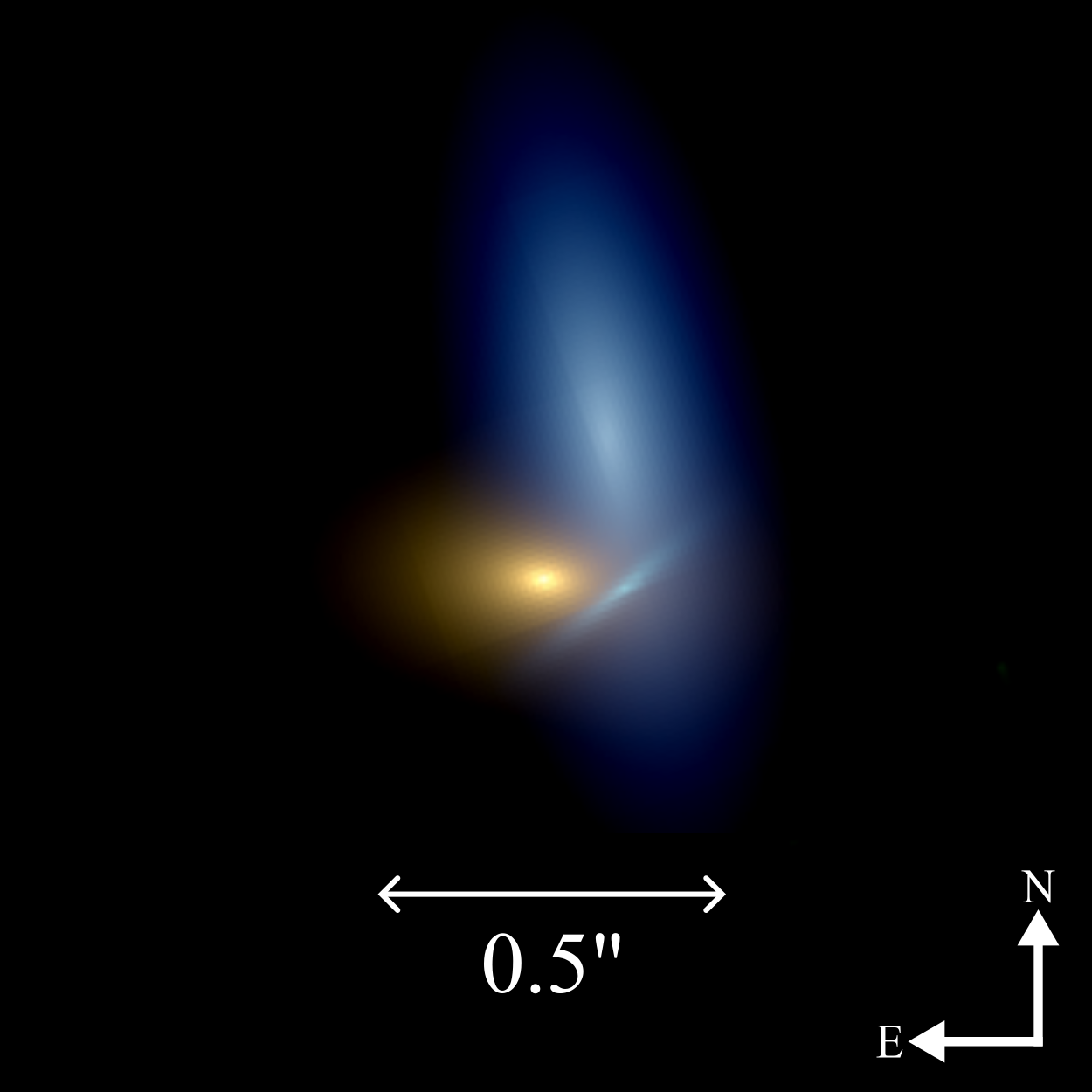}     
 \caption{{\textit Top:} Grey-scale rendering of the \slfit{} reconstructed background source morphology in the F277W band. Caustic lines are overlaid in red. {\textit Bottom:} False color representation obtained by combining reconstructions of the background source in the F444W (red channel), F277W (green channel), and F150W (blue channel) bands. North is up and East is left.
 }
  \label{fig:slfit_src}
\end{figure}

We model the whole system using the forward light profile lensing fitting code \slfit{}. This code has been extensively used in the literature \citep[e.g.][]{Gavazzi2008, Gavazzi2011, Gavazzi2012, brault_extensive_2015,Yang2019} and allows one to fit multiple analytical profiles to the source and the lens. 
We readily assume the total mass distribution to be described as a singular isothermal ellipsoid (SIE), arguably the simplest possible mass distribution able to give reliable estimates of the Einstein radius and source intrinsic flux. 
Its three free parameters are its velocity dispersion, ellipticity, and position angle on the plane of the sky. The Einstein radius $\theta_{\rm Ein}$  and velocity dispersion $\sigma_{\rm SIE}$ are related through the relation
\begin{equation}
\theta_{\rm Ein} = 4 \pi \left(\frac{\sigma_{\rm SIE}}{c}\right)^2 \frac{d_{\rm ls}}{d_{\rm s}},
    \label{eq: rein_def}
\end{equation}
with $\theta_{\rm Ein}$ in radian, $c$ the speed of light, and $d_{\rm s}$ and $d_{\rm ls}$ the angular diameter distances between the observer and the source and between the lens and the source, respectively.
We force the center of the total matter distribution to match the center of the foreground light distribution, i.e. the main lens galaxy. Besides ellipticity and orientation, $\theta_{\rm Ein} $ is the third parameter describing the mass model. Only when attempting to convert this angular scale into a mass (or velocity dispersion) does one need to know the redshifts of the lens and the source in order to compute $d_{\rm ls}/d_{\rm s}$. Constraints on $\theta_{\rm Ein}$ are therefore independent of redshift. On the other hand, the mass enclosed within the Einstein radius is
\begin{equation}
    M(<\theta_{\rm Ein}) = \frac{c^2}{4 G} \frac{d_{\rm l} d_{\rm s}}{d_{\rm ls}} \theta_{\rm Ein}^2 \;,
\end{equation}
with $G$ the constant of gravitation and $d_{\rm l}$ the angular diameter distance between the observer and the lens, so it does depend on angular diameter distances. Indeed, assuming $z_{\rm lens} = 2$ and that the source is located at $z_{\rm source} = 5.5$ rather than $z_{\rm source} = 3$, this would lower the estimate of $M(<\theta_{\rm Ein})$ at  $z_{\rm source} = 3$ by roughly 50\%. 


As anticipated in the previous section, the complex ring's morphology is well resolved and hinders a fit of a simple Sérsic source. We found that 3 components successfully capture the ring's structure in all \NIRCAM{} bands without obvious under-fitting with significant residuals at the position of the ring. However, in order to control the fit with many degrees of freedom, we proceed in an iterative way by
\begin{enumerate}[label=(\roman*)]
    \item fitting the deflector's light while masking the ring.
    \item Keeping the lens unchanged while fitting the bright South West satellite emission in all \NIRCAM{} bands.
    \item Keeping both the lens and satellite unchanged while fitting the first red component (hereafter Comp-1) along with the SIE mass model.
    \item Fitting again the deflector's light while keeping the Comp-1 unchanged.
    \item Fixing the lens and the mass model and fitting for the brightest (and most extended) blue second component (hereafter Comp-2).
    \item While fixing the lens and constraining the range of model parameters for Comp-1 and Comp-2, we update the mass model along with the model for the background source.
    \item Keeping all of the above fixed, we add a third blue component.
    \item Finally, all the parameters describing the mass and the lensed source are fitted together assuming that the previous steps allowed to prevent any significant leftover coupling between the foreground light and the ring photometry.
\end{enumerate}
Throughout this process, we fixed the Sérsic index of both the lens and potential satellite to $n=3$ and we also assumed $n=1$ for the lensed background components. By doing so, we found that we were able to accurately fit the light distribution of the lens, companion, and background source while avoiding introducing additional free parameters that could add degeneracies, in particular between the Sérsic index and the effective radius of the lens \citep[e.g. see the discussion in][]{Graham1996}. As illustrated in Fig.\,\ref{fig:Observations/morphology}, the morphology of the lens is equally well fitted with a free (\SX{}{}, rightmost column) or fixed (\slfit{}, third column from the left) Sérsic index, meaning the total fluxes are equally well recovered in both cases.

When additional source plane components are considered, they are first added as circular profiles, and then their azimuthal structure is allowed to vary.
The parameter space is explored with 16 parallel Monte-Carlo Markov Chains with a burn-in phase. The intermediate chains are relatively short but large excursions are allowed from time to time in order to avoid the chains from falling into local minima. The photometry and mass model parameters and uncertainties are only derived from the MCMC samples of the last fit described in point (viii).

Photometry for the other bands (all but \NIRCAM{}) is derived keeping the morphology of the sources (either lensed or not) constant but fitting only their total flux, accounting for the appropriate PSF in each band. It is worth noting here that we assume a positivity prior on the flux in all the bands which may increase the recovered fluxes of very low signal-to-noise bands compared to more standard aperture photometry.

\subsubsection{Results}

Figure \ref{fig:slfit_src} shows the best fit reconstructed source plane model as found by \slfit{} in the F277W band. A false color image combining F444W (red channel), F277W (green channel), and F150W (blue channel) is also shown. We clearly see the three Sérsic components that were fitted. One of the components is red and compact. It is located East of the caustic and produces the clumps \ClumpWest{} and \ClumpEast{} in the image plane. The other two blues components produce the bluer part of the ring.

The main parameters of the lens' mass distribution are given in Table\,\ref{tab:sec:Lens model/mass distribution}. This includes the Einstein radius $\theta_{\rm{Ein}}$, magnification $\mu$, and total mass measured within $\theta_{\rm{Ein}}$. For the latter, we include three different values obtained using the three redshift solutions derived by \Lephare{}, \Cigale{}, and \EAZY{} on \slfit{} photometry and $z_{\rm lens} = 2.00 \pm 0.02$. We stress here that magnification depends heavily on the choice of the mass distribution and in particular on its inner slope (here SIE implies $\gamma \equiv -d \log \rho / d\log r = 2$). The magnification that is given corresponds to a mean value derived as the flux weighted sum of the magnification experienced by the three components of the background source.

\subsection{Source reconstruction And Density Slope using \pyautolens{}}
\label{sec:Lens model/PyAutoLens}

\subsubsection{Method}

\begin{figure*}
    \centering
    \includegraphics[width=0.24\textwidth]{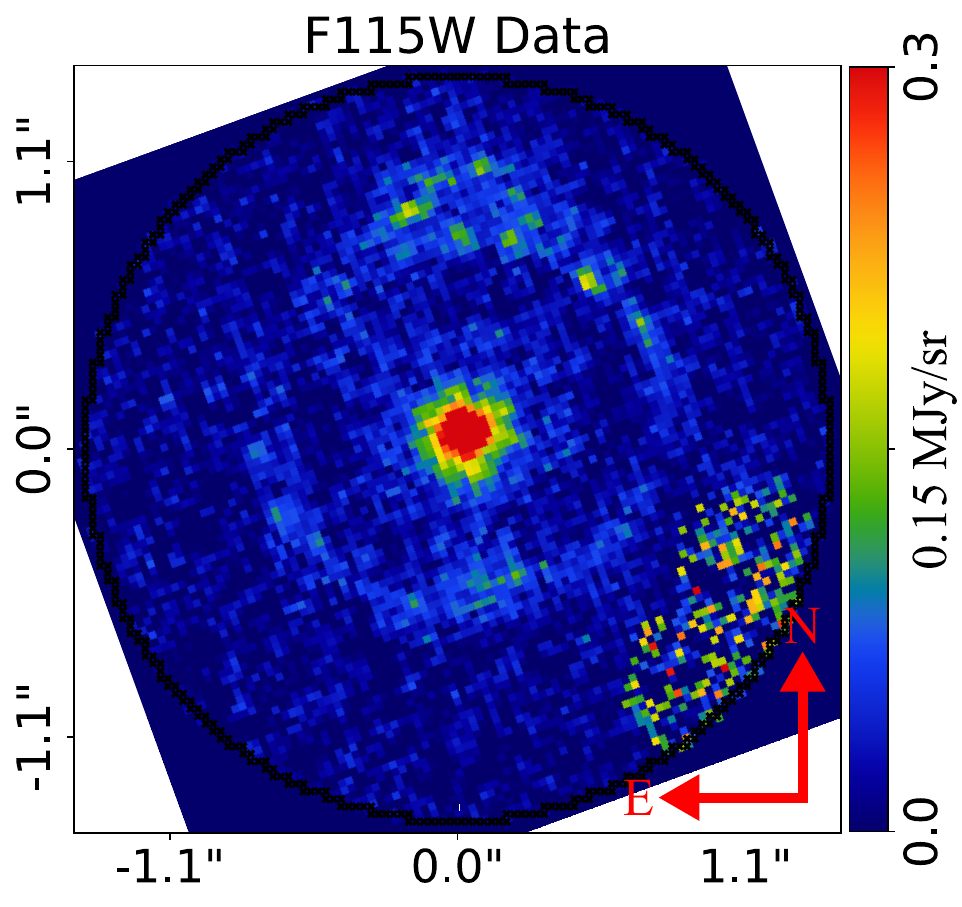}
    \includegraphics[width=0.24\textwidth]{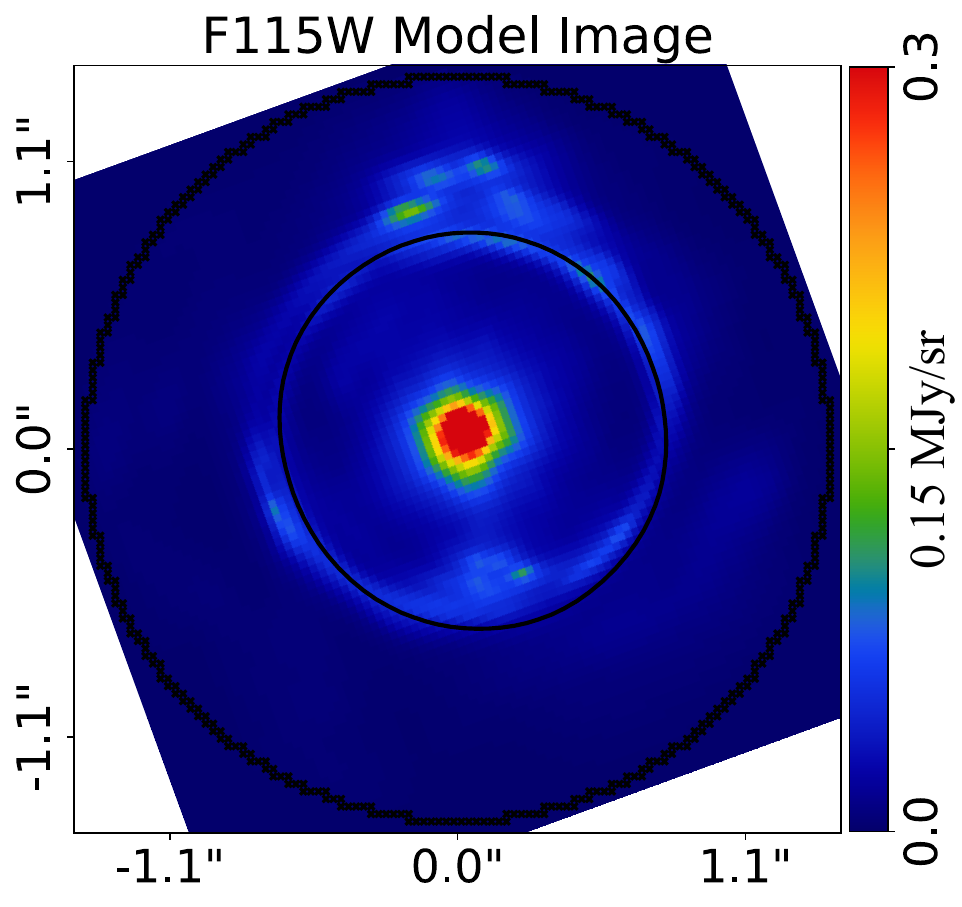}
    \includegraphics[width=0.24\textwidth]{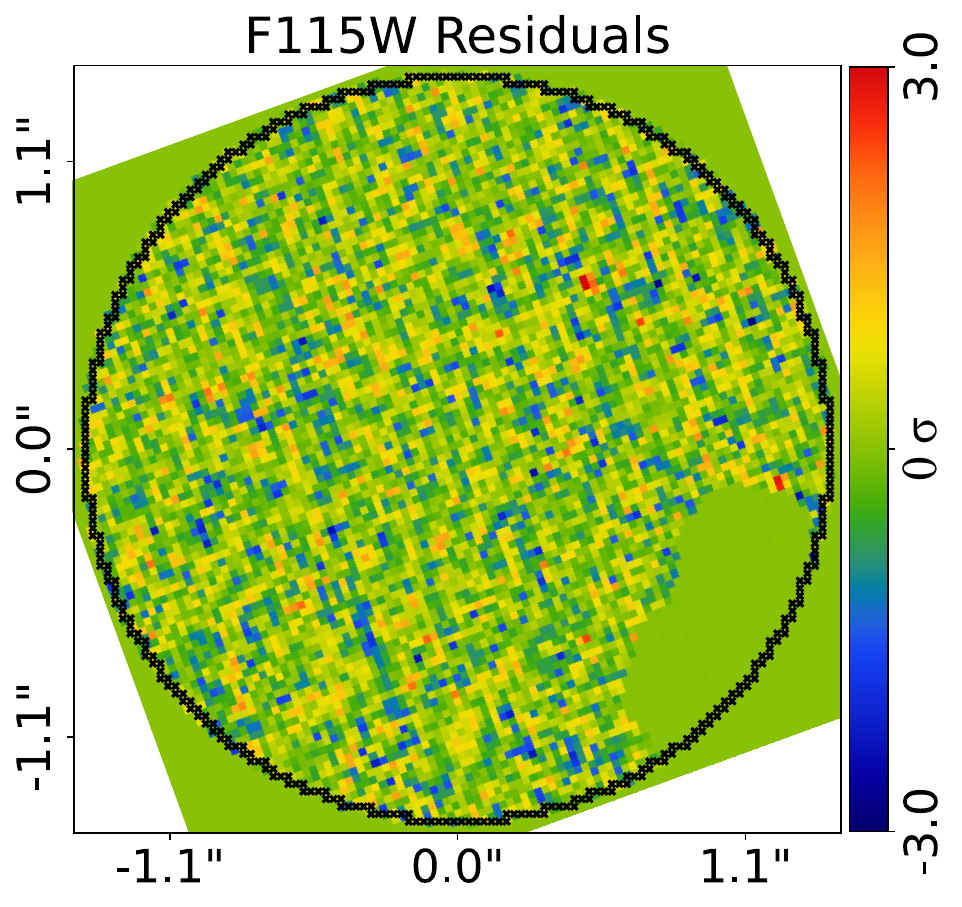}
    \includegraphics[width=0.24\textwidth]{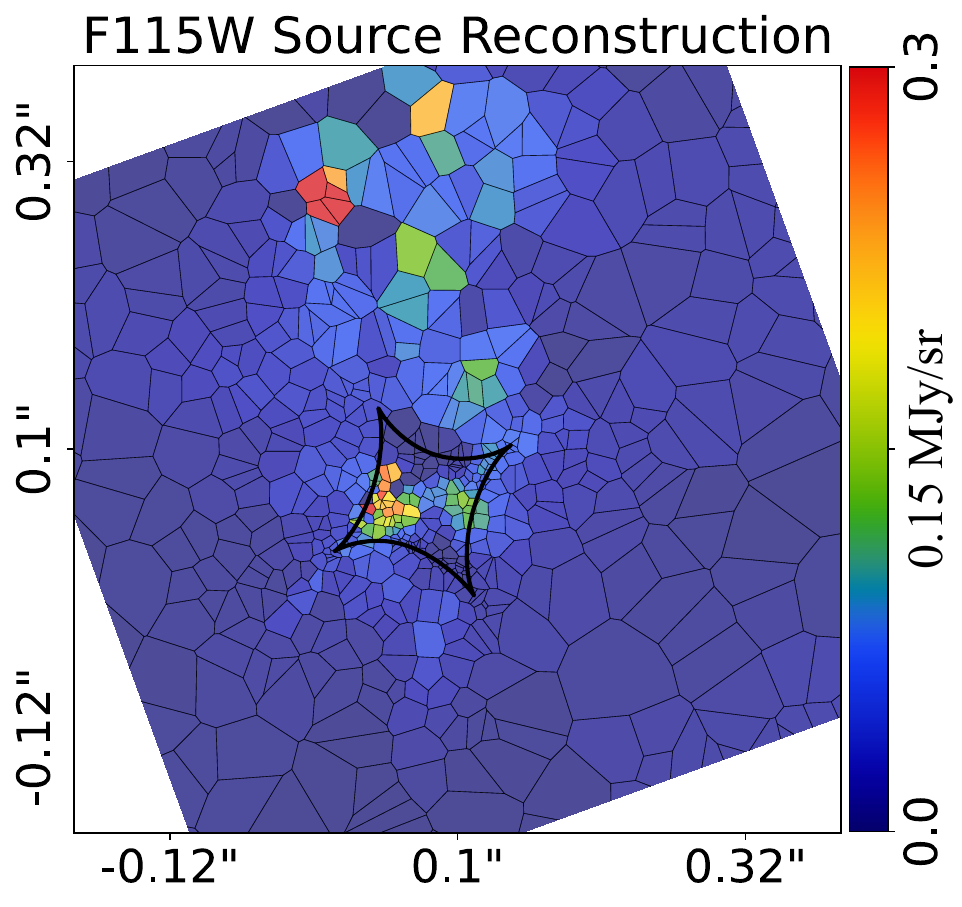}
    \includegraphics[width=0.24\textwidth]{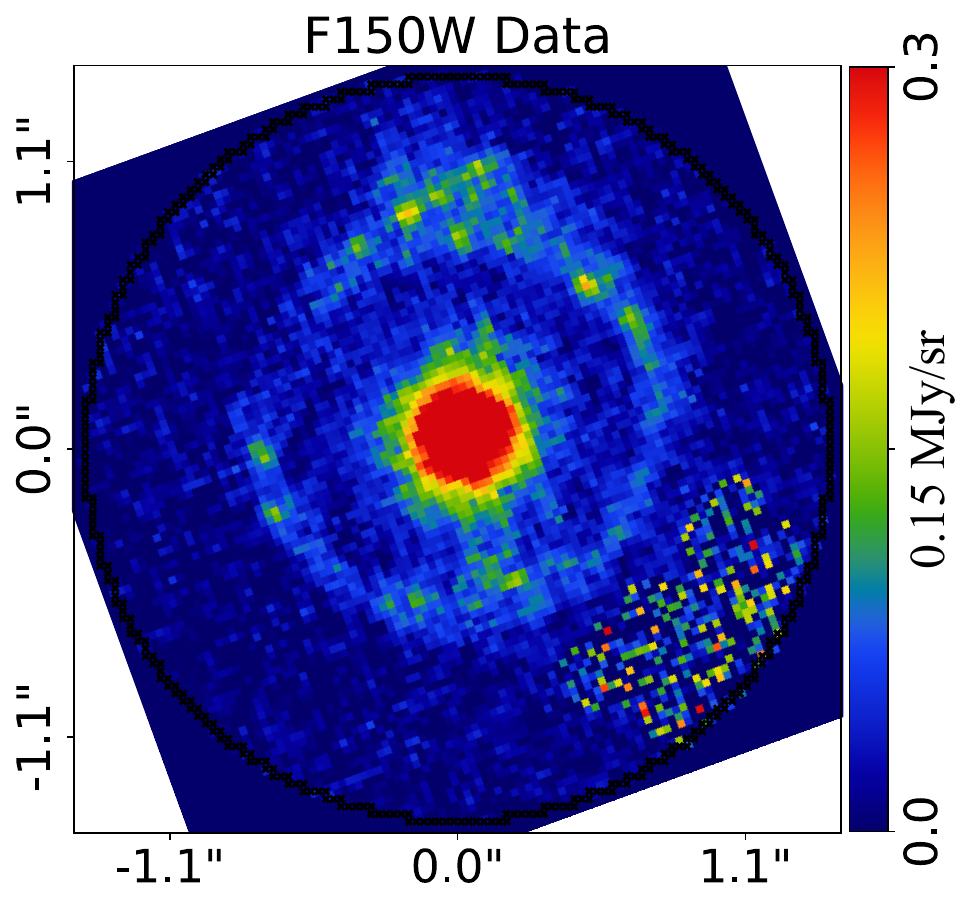}
    \includegraphics[width=0.24\textwidth]{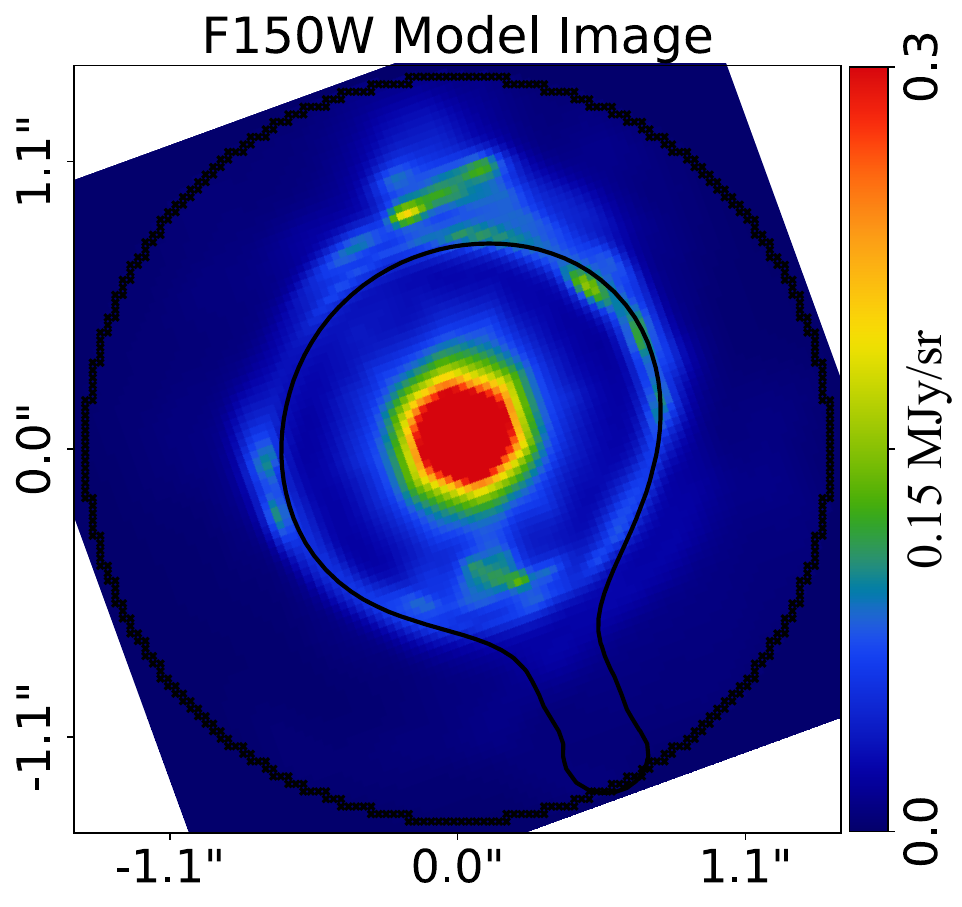}
    \includegraphics[width=0.24\textwidth]{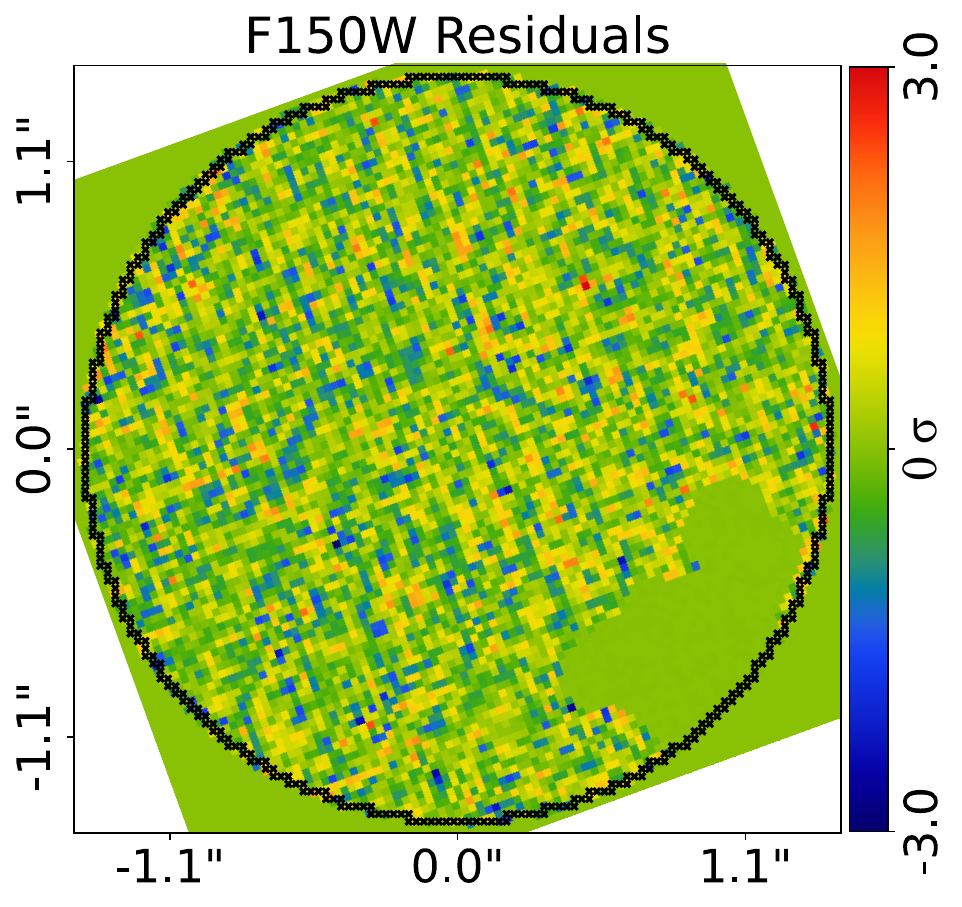}
    \includegraphics[width=0.24\textwidth]{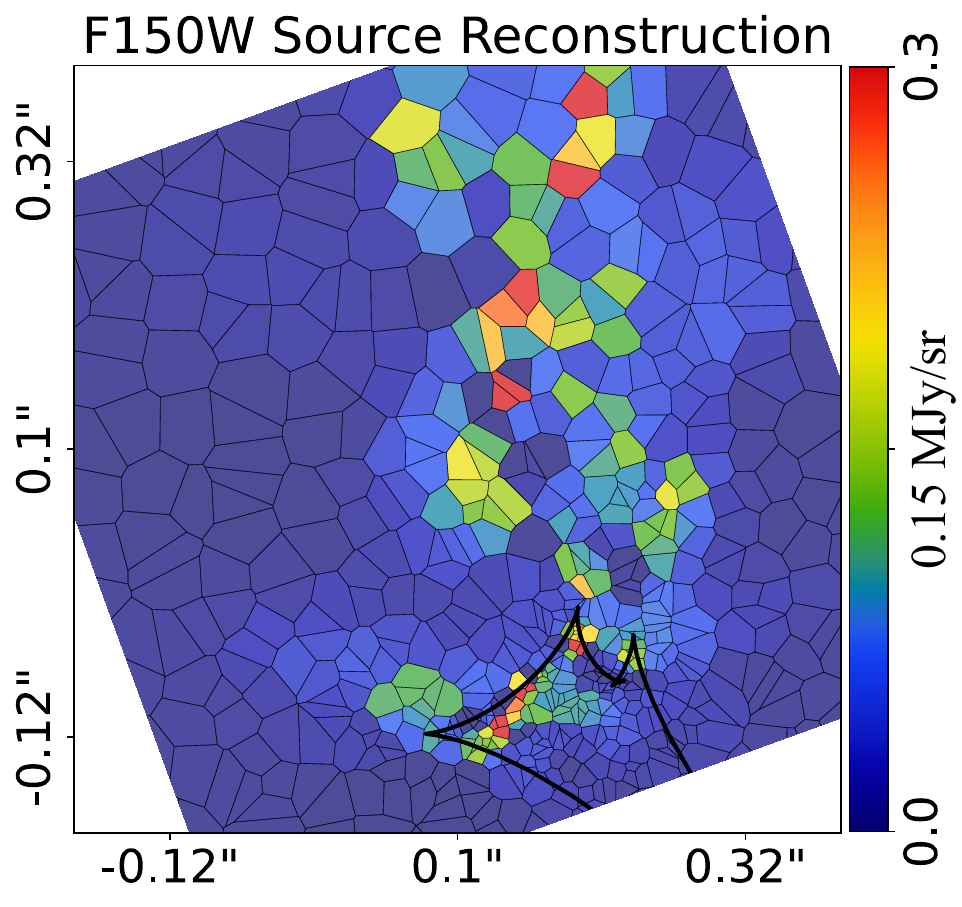}
    \includegraphics[width=0.24\textwidth]{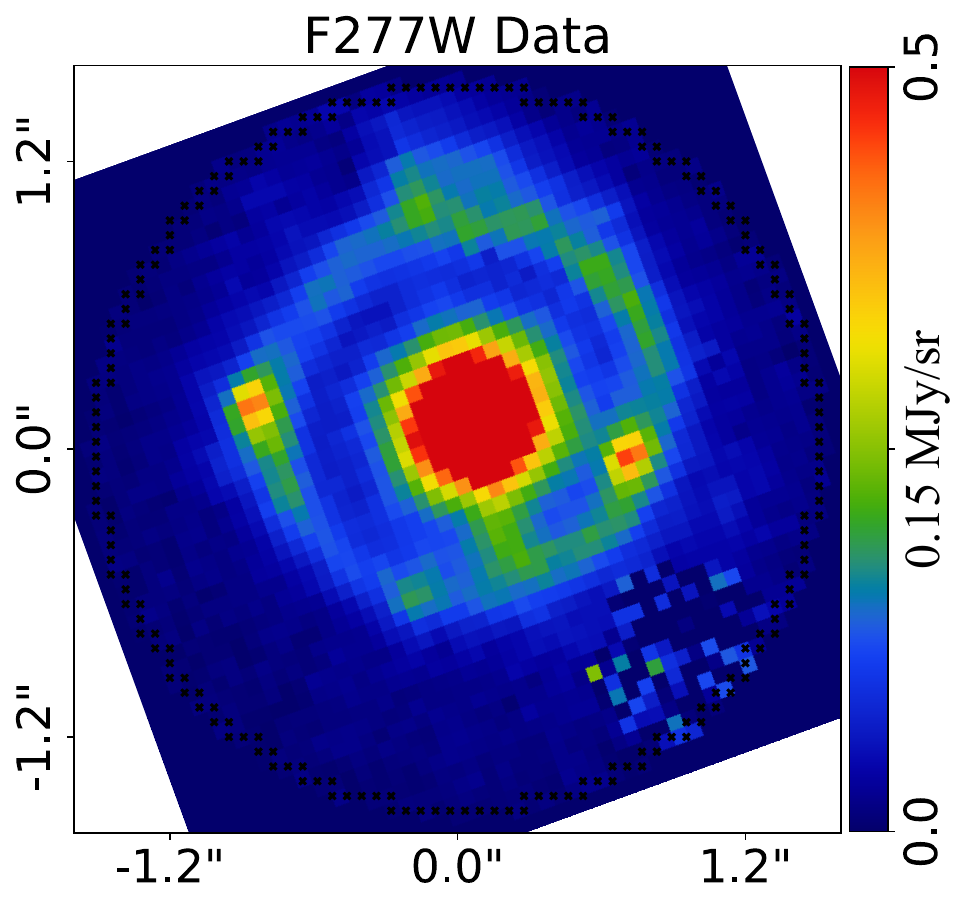}
    \includegraphics[width=0.24\textwidth]{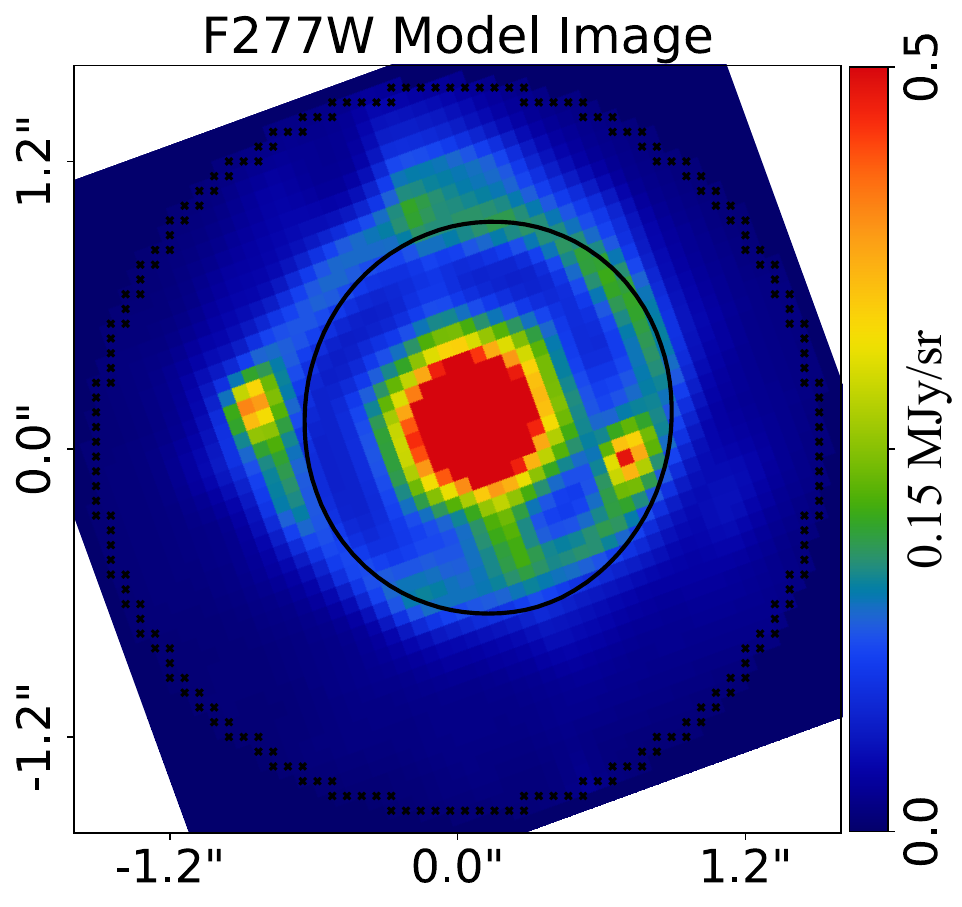}
    \includegraphics[width=0.24\textwidth]{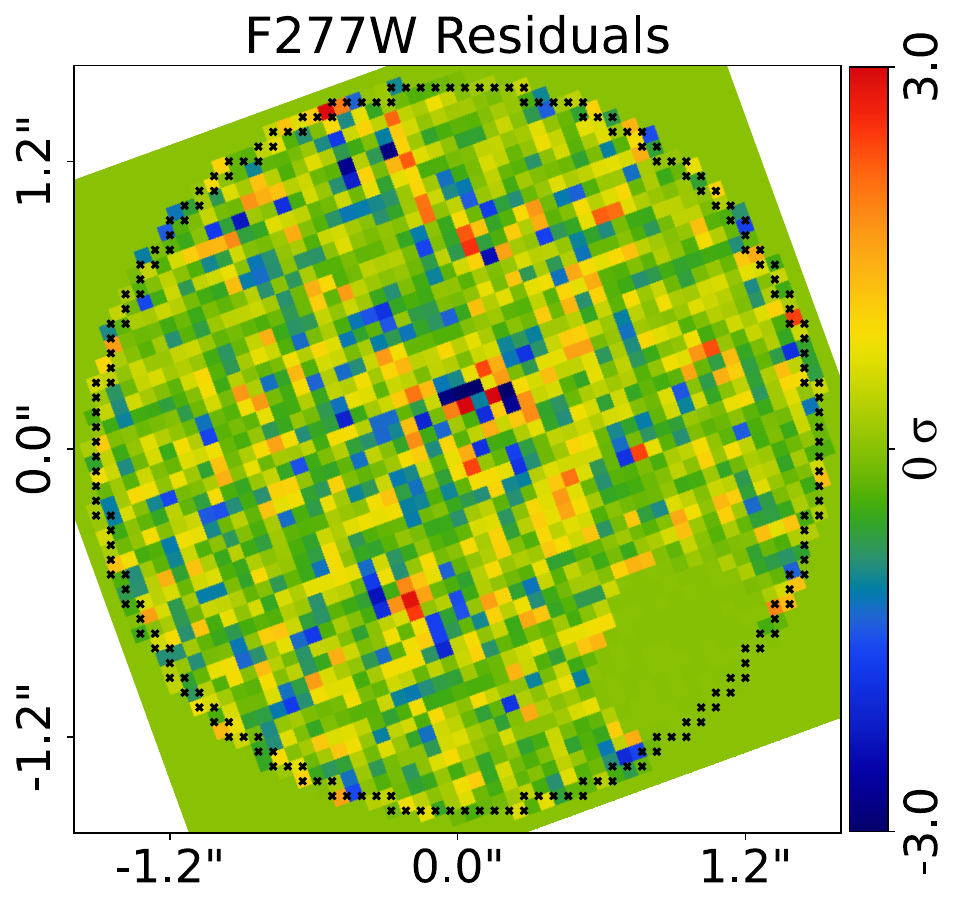}
    \includegraphics[width=0.24\textwidth]{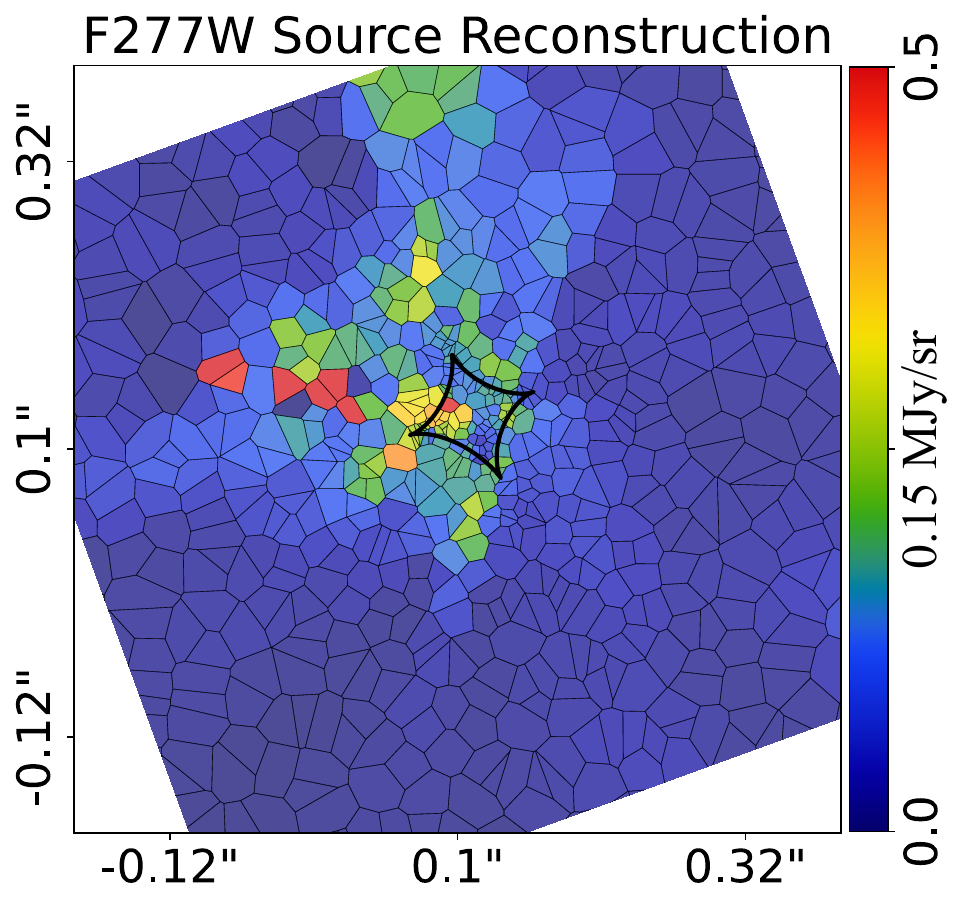}
    \includegraphics[width=0.24\textwidth]{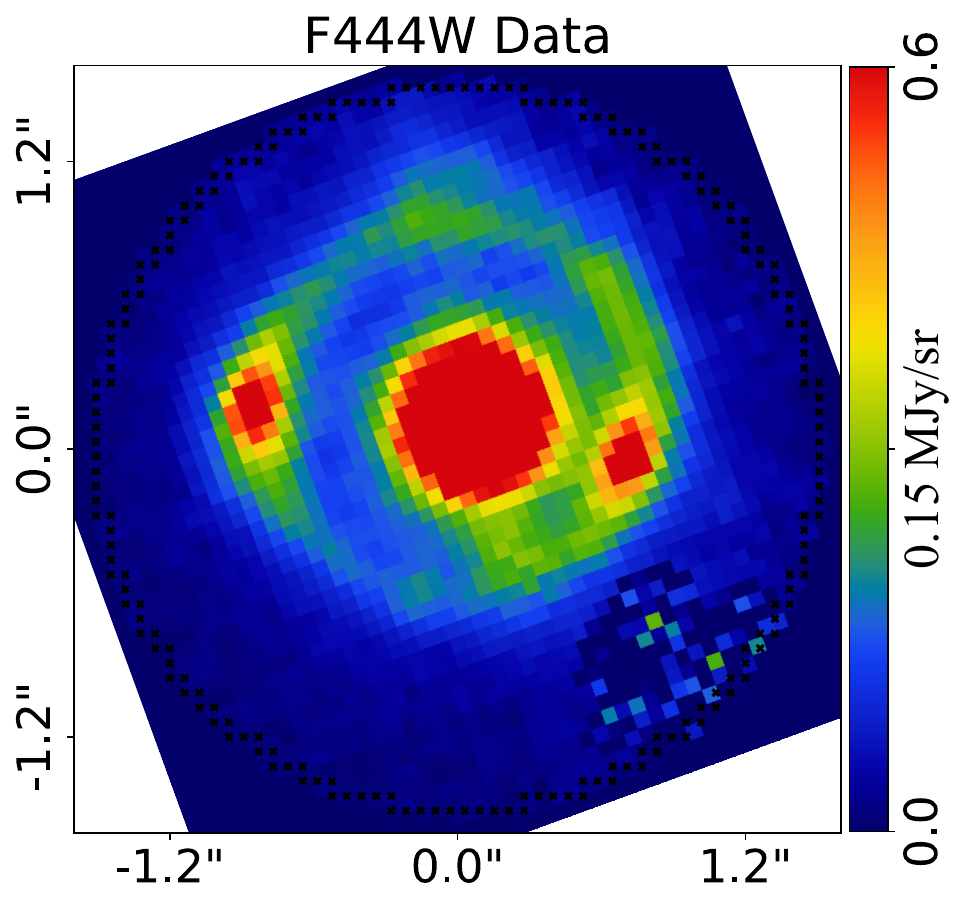}
    \includegraphics[width=0.24\textwidth]{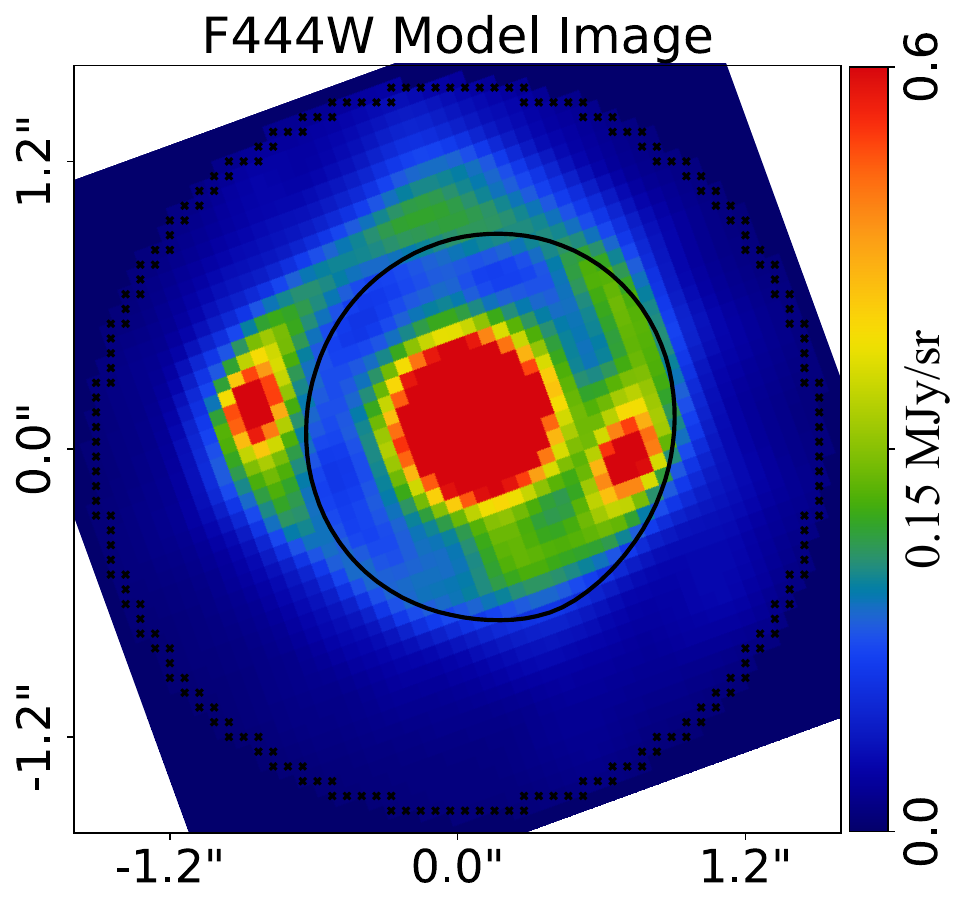}
    \includegraphics[width=0.24\textwidth]{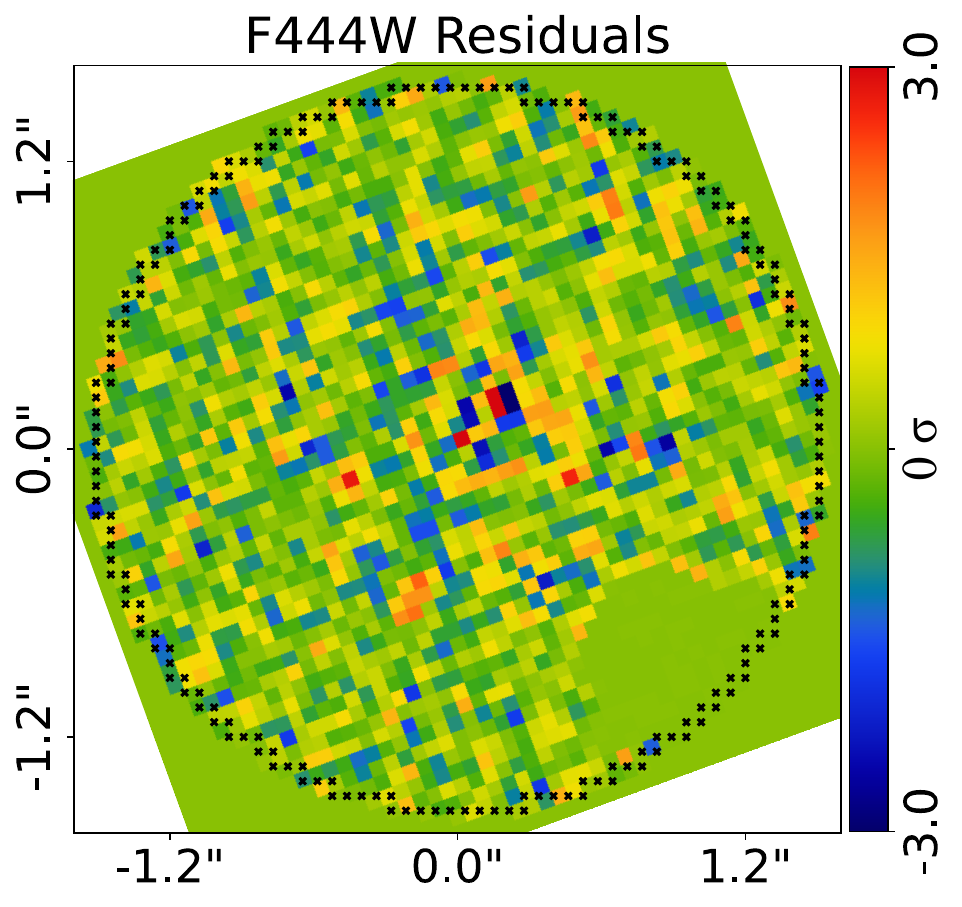}
    \includegraphics[width=0.24\textwidth]{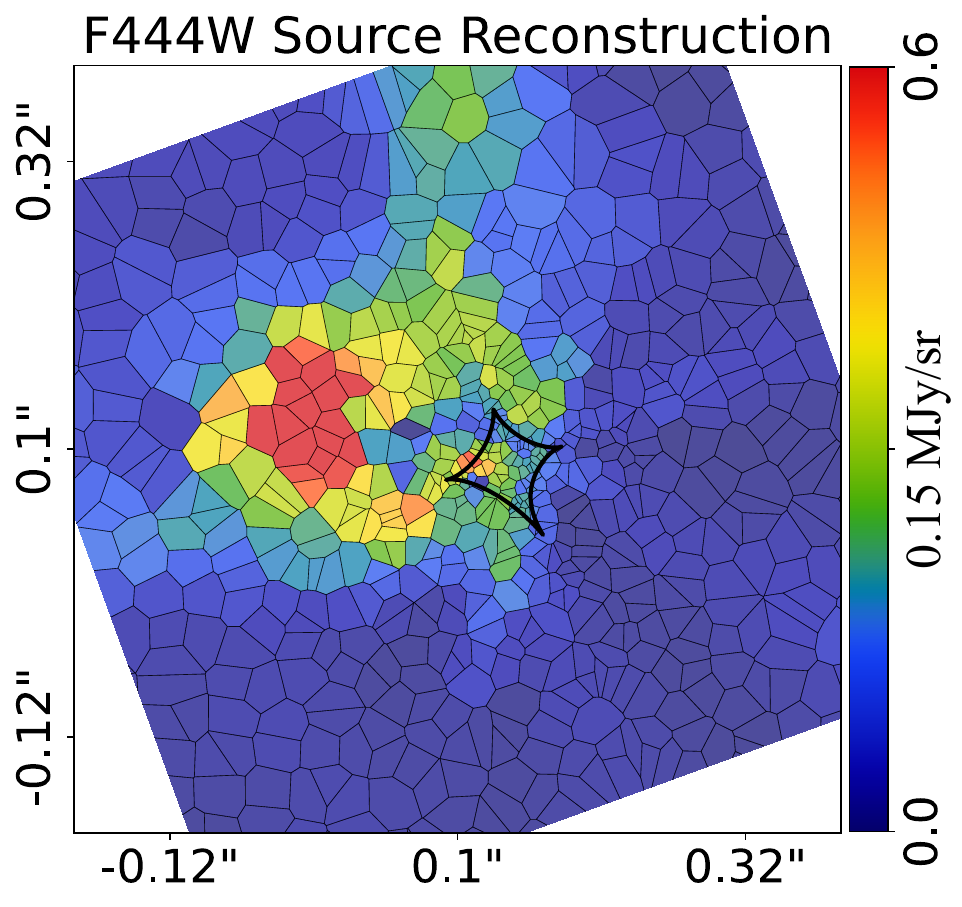}
    \caption{Results of the independent \pyautolens{} lens model fits to the F115W, F150W, F277W and F444W bands. From left to right: observed image, lens model image, normalized residual map and source reconstruction. North is up and East is left.}
    \label{fig:Lens model/PyAutoLens_Fit}
\end{figure*}

Next, we perform lens modeling using the open-source software \pyautolens{}\footnote{\url{https://github.com/Jammy2211/PyAutoLens}} \citep{Nightingale2018, Nightingale2021}. In contrast to \slfit{}, the \pyautolens\ analysis: (i) reconstructs the unlensed source galaxy on a Voronoi mesh which can account for irregular and asymmetric features \cite[e.g. mergers,][]{Nightingale2015, Nightingale2018}; and (ii) uses information contained in the lensed source's extended surface brightness distribution to measure more detailed properties of the lens galaxy's mass, in this case, the power-law density slope. We give a concise overview of the aspects of lens modeling that are most important for this study. 

Before lens modeling, preprocessing steps are performed on the data. Each band is modeled using the pixel scale closest to the native scale, that is $0.03"$ for F115W and F150W bands, and $0.06"$ for F227W and F450W bands. A $2.6"$ circular mask is applied to all datasets, defining the region within which the lens analysis is performed. The emission from the feature to the South West of the Einstein ring (see Fig.\,\ref{fig:Observations/color-image}) is removed from each image via a graphical user internal, which replaces the emission with Gaussian noise. Fits including this feature were performed but the lens model indicated there was no lensing counterpart, indicating that it is a foreground galaxy. We use an adaptation of the Source, Light and Mass (SLaM) pipelines described in \cite{Etherington2023} and \cite{Nightingale2023}. Lens models are fitted using the nested sampling algorithm {\tt nautilus} \citep{Lange2023}. These pipelines automate the lens model fitting and are used to fit all four waveband images (F115W, F150W, F227W, F444W) independently.

The foreground lens galaxy's emission is modeled and subtracted using a multiple Gaussian expansion \citep[MGE, ][]{Cappellari2002}. The MGE is implemented internally within \pyautolens{} and performed simultaneously with the source reconstruction (He et al. in prep). The MGE decomposes the lens emission into 100 elliptical two dimensional Gaussians. Their axis-ratios, position angles and sizes vary, capturing departures from axisymmetry. The intensity of every Gaussian is solved simultaneously with the source reconstruction, using a non-negative least square solver (NNLS). The MGE provides a clean deblending of lens and source light.

The source is reconstructed using an adaptive Voronoi mesh with 2000 pixels for the higher resolution F115W and F150W bands and 1600 pixels for the F277W and F444W bands. The Voronoi pixel distribution adapts to the source morphology. This uses the natural neighbour interpolation and adaptive regularization described in \cite{Nightingale2023}, but unlike this study enforces positivity on source pixels by using the NNLS. The Voronoi mesh is able to reconstruct irregular galaxy morphologies, provided the mass model is accurate. 

The lens galaxy mass model is used to ray-trace light from the image-plane to the unlensed source-plane, where the source reconstruction is performed. We fit an elliptical power-law (PL) mass distribution (representing the stars and dark matter) with convergence \citep{Suyu2012}
\begin{equation}
    \kappa(x,y) =
    \frac{\Sigma(x,y)}{\Sigma_\mathrm{crit}} =
    \frac{3-\gamma}{1+q}\bigg(\frac{b}{\sqrt{x^2+y^2/q^2}}\bigg)^{\gamma-1},
    \label{eq: kappa}
\end{equation}
where $\Sigma(x,y)$ is the mass density, $\gamma$ is the logarithmic slope of the mass distribution in 3D, $1 \geq q > 0$ is the projected minor to major axis ratio of the elliptical isodensity contours, and $b\geq0$ is the angular scale length of the profile. The special case $\gamma=2$ recovers the SIE mass distribution fitted above, and $q = 1$ recovers the Spherical Isothermal Sphere (SIS). The profile has additional free parameters for the central coordinates $(x_{\rm c}, y_{\rm c})$ and position angle $\phi$, measured counterclockwise from the positive $x$-axis. When varying the ellipticity, we actually sample from and adjust free parameters
\begin{align}
\varepsilon_{1} &=\frac{1-q}{1+q} \sin 2\phi~, &
\varepsilon_{2} &=\frac{1-q}{1+q} \cos 2\phi~,  
\label{eq: ellip}
\end{align}
because these are defined continuously in $-1<\varepsilon_{i}<1$, eliminating the periodic boundaries associated with angle $\phi$ and the discontinuity at $q=0$.  
We similarly parameterise the external lensing shear as components $\gamma_{1\text{ext}}$ and $\gamma_{2\text{ext}}$. The external shear magnitude $\gamma_{\text{ext}}$ and angle $\phi_{\text{ext}}$ are recovered from these parameters by
\begin{align}
    \gamma_{\text{ext}} &= \sqrt{\gamma_{1\text{ext}}^2+\gamma_{2\text{ext}}^2}~, &
    \tan{2\phi_{\text{ext}}}&=\frac{\gamma_{2\text{ext}}}{\gamma_{1\text{ext}}}~.
    \label{eq: shear}
\end{align}

We include the foreground galaxy to the South West in the mass model using an SIS profile, where the centre is fixed to the brightest pixel of this galaxy.

\subsubsection{Results}

\begin{figure*}[htbp]
    \centering
    \includegraphics[width=\hsize]{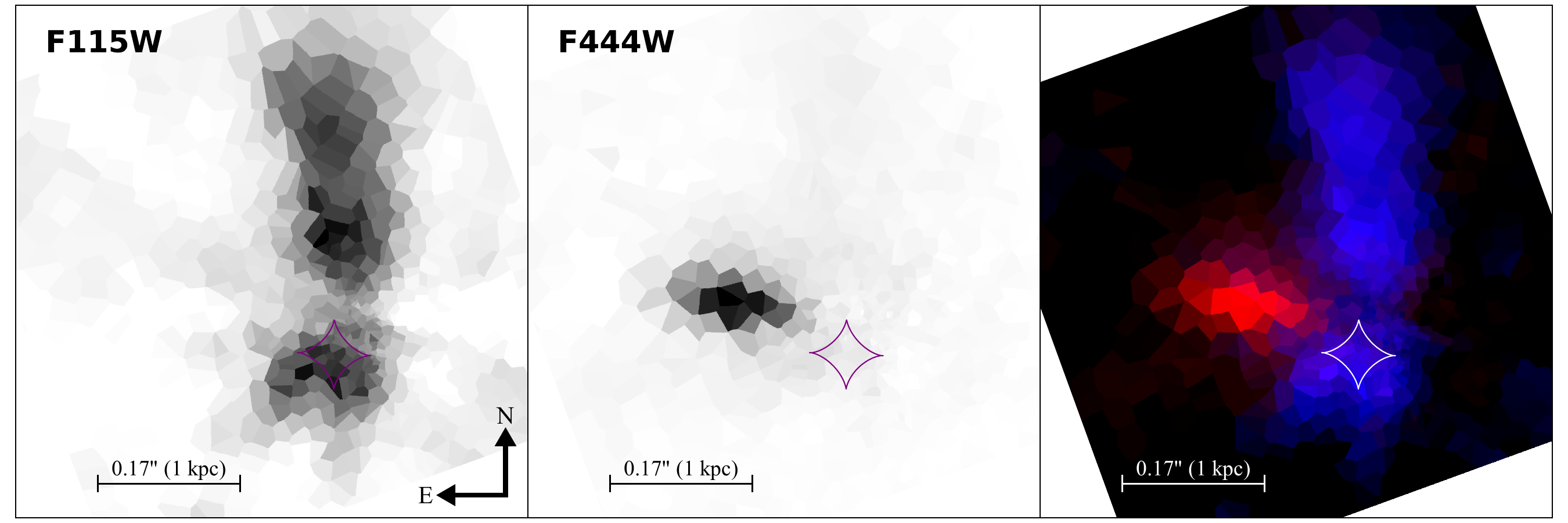}
    \caption{Source reconstruction from \pyautolens{} on \JWST{}/\NIRCAM{} F115W (left-hand plot) and F444W (middle plot) bands using identical mass models. The right-hand plot represents the superposition of the two reconstructions with F115W in blue and F444W in red. In each panel, the caustic lines are shown. North is up and East is left.}%
    \label{fig:Lens model/PyAutoLens}
\end{figure*}

Figure \ref{fig:Lens model/PyAutoLens_Fit} shows the observed image, model, normalized residuals, and source reconstruction of fits to the \Cring{}. Across all four \JWST{} wavelengths, the foreground lens and lensed source emission are fitted accurately, as visible in the residuals of Fig.\,\ref{fig:Lens model/PyAutoLens_Fit}. The overlaid black lines show the tangential critical curves and caustics. They are similar across each wavelength, indicating that the mass models are generally consistent. The Voronoi source reconstructions show the striking change in appearance of the source galaxy across wavelengths, where F115W and F150W filters reveal clumpy and elongated emission from North to South and F277W and F444W filters reconstruct an offset and compact component oriented from East to West and that is not visible at bluer wavelengths. We note that this source reconstruction is broadly consistent with that of \slfit{} that find three Sérsic components whose spatial offset and spatial extent match the reconstructed morphology of \pyautolens{}. We discuss in Sect.\,\ref{sec:Properties/source/dust} the potential meaning of these two different components.

The reconstructions in Figure \ref{fig:Lens model/PyAutoLens_Fit} indicate the source's red compact feature is not coincident with the clumpy emission reconstructed at bluer wavelengths. To confirm this, we reconstruct all wavebands using a single unified mass model derived from the highest S/N F444W image, which is shown in Fig.\,\ref{fig:Lens model/PyAutoLens}. The emission in F444W is clearly offset from that in F115W and located in a distinct region of the source-plane. Different wavelength observations are therefore detecting different components of the lensed source galaxy with the emission in F444W associated to the optical (around \SI{6700}{\angstrom}) and the emission in F115W associated to the UV (around \SI{1800}{\angstrom}).

The inner slope of the total mass profile inferred for the F115W, F150W, F277W and F444W filters are: ${2.78}^{+0.20}_{-0.36}$, ${2.62}^{+0.32}_{-0.32}$, ${2.16}^{+0.12}_{-0.20}$ and ${2.12}^{+0.10}_{-0.17}$. The bluer F115W and F150W slope values are consistent with one another, as are the redder F277W and F444W values. However, the results at blue and red wavelengths appear inconsistent. These distinct slope measurements correlate with the bluer filters having a distinct source galaxy from the redder filters. We discuss this result more in Sect.\,\ref{sec:Properties/lens/slope}.

\section{Discussion}
\label{sec:Properties}

\subsection{Mass budget of the central lens}
\label{sec:Properties/lens}


\begin{figure*}[htbp]
  \centering
  \begin{subfigure}[b]{0.475\textwidth}
            \includegraphics[width=1\hsize]
            {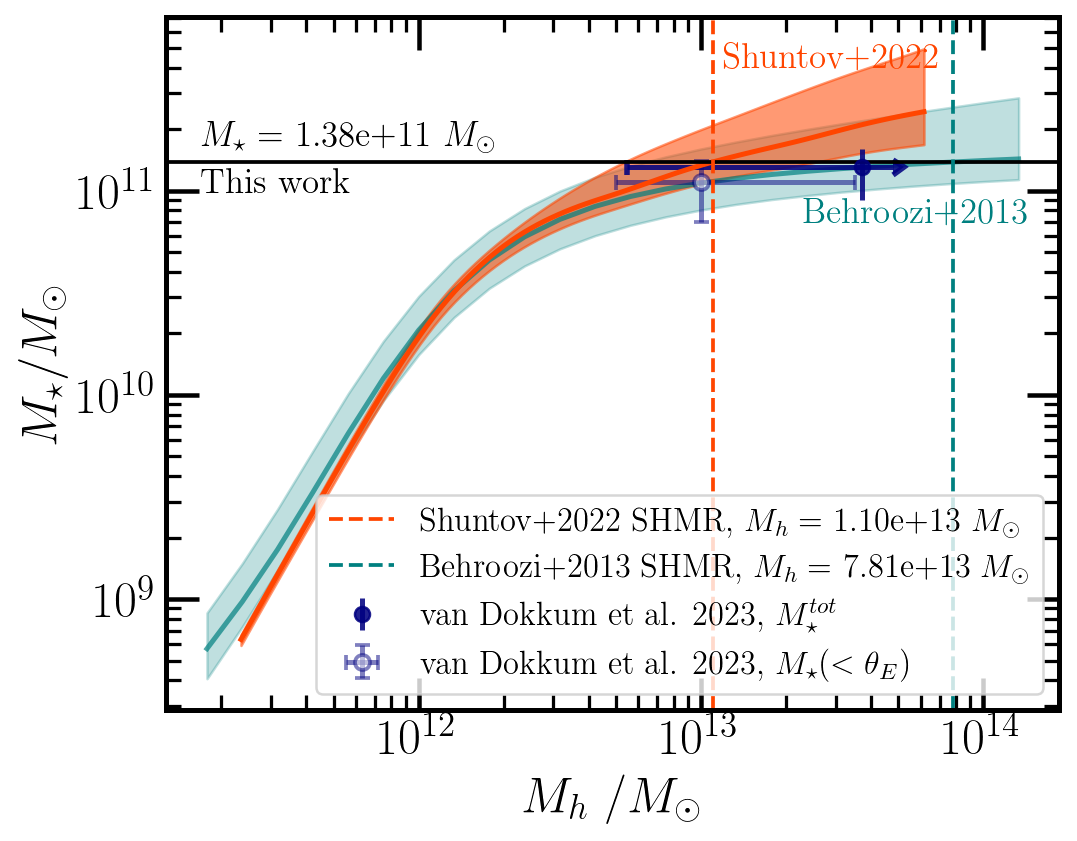}     
  \end{subfigure}
  \begin{subfigure}[b]{0.49\textwidth}
            \includegraphics[width=1\hsize]{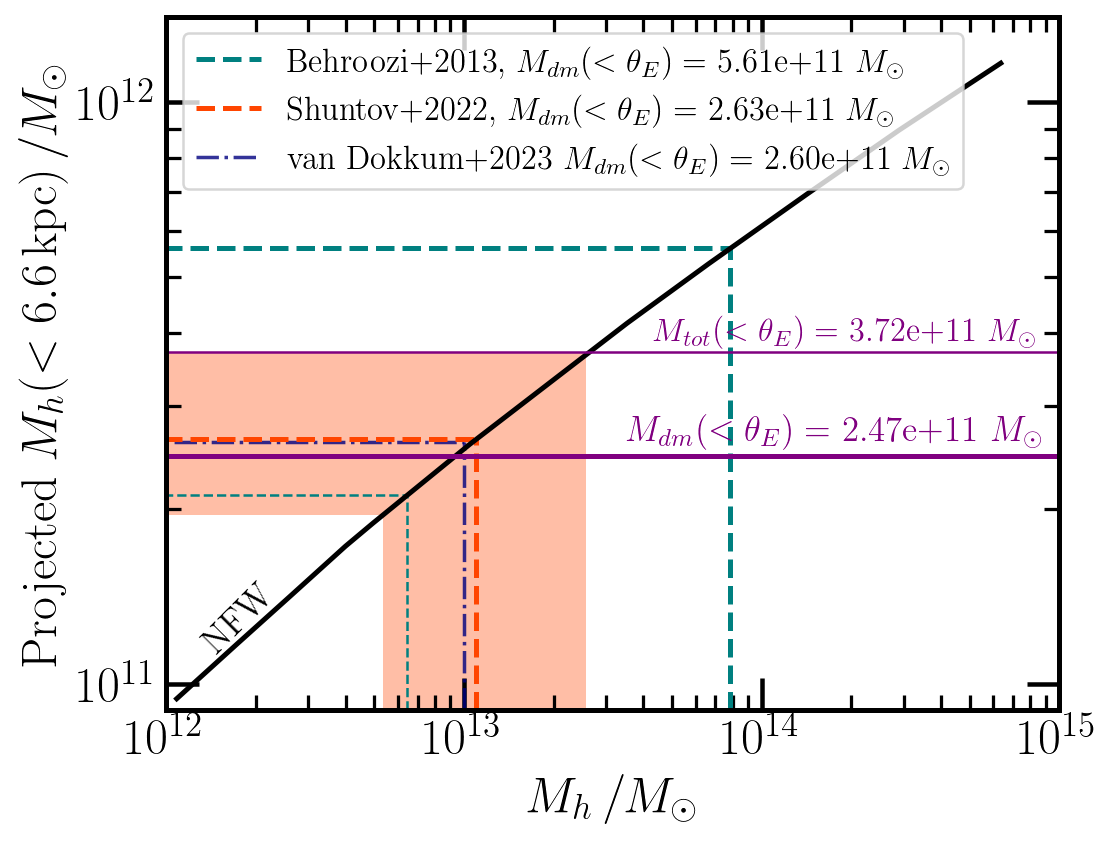}     
  \end{subfigure}
 \caption{{\bf Left:} Stellar-to-halo mass relations (SHMR) from \cite{Shuntov2022} (in orange) and \cite{Behroozi13} (in teal) at $z\sim2$ that we use to estimate the dark matter halo mass in which the lens resides. The solid black horizontal line marks the total stellar mass that we estimate for the lens, while the dashed vertical orange and teal lines mark the halo mass obtained from \cite{Shuntov2022} and \cite{Behroozi13} SHMR respectively. For comparison, we also show the SHMR points from \cite{vandokkum2023massive} that we derive using their total stellar mass and the \cite{Behroozi13} SHMR as shown in the solid blue marker. The transparent blue marker shows the stellar and halo masses quoted in \citet{vandokkum2023massive}. {\bf Right:} Relation between projected mass within the Einstein radius and total halo mass. The orange dashed line and shaded regions show the results from \cite{Shuntov2022} SHMR, while the teal thick and thin dashed lines show results and the lower limit on the uncertainty from \cite{Behroozi13} SHMR. The purple thick line shows the dark matter mass derived as ${M}_{\rm tot}(<\theta_{\rm Ein})-{M}_{\star}(<\theta_{\rm Ein})$, while the thin one shows the total mass derived from the \textsc{sl\_fit} modeling. For comparison, we also show the results quoted in \cite{vandokkum2023massive} in a dashed-dotted blue line.
 }
  \label{fig:dark-matter-content}
\end{figure*}

We first discuss the central lens which is among the most massive galaxies at $z\sim 2$ given that it lies between 0.2 and \SI{0.6}{\dex} above the characteristic stellar mass \citep{Schechter76} of the total and passive stellar mass functions derived by \citet{Weaver2022cosmos2020}. Its star-formation is quenched as shown by rest-frame NUV-r/r-K colors that are characteristic of evolved and passive galaxies \citep[e.g.][]{Arnouts2013,Moutard2020b}, and by its sSFR which we find to be of the order or below \SI{e-13}{\per\year} (see Table~\ref{tab:Redshift/lens}). Its morphology is consistent with that of an elliptical, well described by a smooth surface brightness profile following a S\'ersic law with a Sérsic index of $n=6.4$ when let to vary as a free parameter. Furthermore, the lens is compact with $R_{\rm eff} = \SI{1.5}{\kilo\pc}$ when fitting with \sextractor{} ($R_{\rm eff} = \SI{2.5}{\kpc}$ with \slfit{}\footnote{While a factor of about 1.7 on the effective radius may appear significant, we note that this parameter is found to be degenerate with the Sérsic index. As illustrated in Fig.\,\ref{fig:Observations/morphology} both models fit equally well the lens.}). This size measurement is consistent with the mass-size relation of passive galaxies found at $z = 2$ \citep[e.g.][]{VanDerWel14}. Given that lensing provides a constraint on the total mass of the lens (dark + baryonic; see \ref{sec:Lens model}) and multi-wavelength observations provide constraints on its stellar mass, the combination of the two can tell us about the DM content of ETGs and how likely they could host hidden gas reservoirs. In turn, this can be used to better understand whether the quenching of the star-formation of ETGs could be the result of in-situ gas stabilization processes \citep[i.e. morphological quenching, e.g.][]{Martig09} or of other mechanisms \citep[for a review on the topic, see][]{Man2018, Moutard2020b}.

First, we compare the stellar and total masses within the Einstein radius $\theta_{\rm Ein} = \SI{0.78(0.04)}{\arcsec}$ which is equal to \SI{6.6}{\kilo\pc} at $z = 2.00$. Using the structural parameters of the lens derived from \slfit{}, we find that 91\% of the total light is encompassed within the Einstein radius ($\theta_{\rm Ein}$). Assuming the mass distribution follows the light distribution, it corresponds to a stellar mass within $\theta_{\rm Ein}$ of $M_\star (< \theta_{\rm Ein}) = 1.25^{+0.13}_{-0.10} \times 10^{11} \unit{\Msun}$. Given the total mass of the deflector within the Einstein radius (see Table~\ref{tab:sec:Lens model/mass distribution}), we conclude that the stellar populations contribute to $(34 \pm 5)$\% of the total mass within $\theta_{\rm Ein}$ and that there is a remaining $\Delta M = M_{\rm tot} (< \theta_{\rm Ein}) - M_{\star} (< \theta_{\rm Ein}) = (2.46 \pm 0.30) \times 10^{11} \unit{\Msun}$ in the mass budget. 

In order to understand if this remaining mass $\Delta M$ could be entirely attributed to the dark matter (DM) halo within the Einstein ring or not, we derive the expected DM halo mass for the lens using the stellar-to-halo mass relation (SHMR) from \citet{Shuntov2022}. This SHMR was derived from a Halo Occupation Model constrained using clustering and stellar mass function measurements from the COSMOS2020 catalogue \citep{Weaver2022cosmos2020}. This SHMR is shown in orange on the left panel of Fig.~\ref{fig:dark-matter-content} and is compared to the SHRM from \citet[shown in teal]{Behroozi13} that was used in the analysis of \citet{vandokkum2023massive}. Using the total stellar mass of the lens derived by \Lephare{} ($M_\star = 1.37^{+0.14}_{-0.11} \times \SI{e11}{\Msun}$), we estimate an expected DM halo mass within its virial radius of $M_{\rm h} = 1.09^{+1.46}_{-0.57} \times 10^{13} \unit{\Msun}$. We then calculate the DM mass encompassed within the Einstein radius by integrating an NFW \citep{nfw} profile\footnote{Where we assume a \cite{DuttonMaccio2014} mass-concentration relationship and not take into consideration the scatter in this relationship in our discussion.} within a cylinder of radius $\theta_{\rm Ein} = \SI{6.6}{\kilo\pc}$. We obtain $M_{\rm h}(<\theta_{\rm Ein}) = 2.63^{+1.08}_{-0.68} \times \SI{e11}{\Msun}$ which is consistent with the $\Delta M = (2.46 \pm 0.30) \times \SI{e11}{\Msun}$ remaining mass not accounted for by the stellar content of the lens. 
Using the SHMR from \citet{Behroozi13} instead would increase the predicted DM halo mass within the Einstein radius by roughly a factor of two. However this SHRM behaves exponentially in this mass regime, making the DM halo mass prediction so uncertain it is effectively consistent with our value derived using the SHRM from \citet{Shuntov2022} (see the dashed and dotted teal lines on the right panel of Fig.\,\ref{fig:dark-matter-content}). Furthermore, we also get consistent results when using the stellar mass derived by \Cigale{} or \EAZY{} on \slfit{} photometry.

We note that both our stellar mass derived from SED fitting and our expected DM halo mass derived from the SHRM are consistent with the values from \citet{vandokkum2023massive}. However, in their analysis, the sum of the two components is not sufficient to account for the total mass derived from lensing whereas, in our case, it sufficient. This is because our total mass estimates within the Einstein radius are different. They find $M_{\rm tot} (< \theta_{\rm Ein}) = 6.5^{+3.7}_{-1.5} \times \SI{e11}{\Msun}$ whereas we find $M_{\rm tot} (< \theta_{\rm Ein}) = \SI{3.66(0.36)e11}{\Msun}$ instead. Since our Einstein radius and lens photometric redshifts agree, the main reason for this large difference is the fact that they find the background source at $z_{\rm{source}} = 2.98^{+0.42}_{-0.47}$ whereas we find it at $z_{\rm{source}} = 5.48 \pm 0.06$. 

Given that the solution at $z < 3$ is disfavored by our SED fitting results when taking into account ground-based observations, we conclude that the mass budget of the lens is consistent with the presence of a DM halo mass of total mass $M_{\rm h} = 1.09^{+1.46}_{-0.57} \times 10^{13} \unit{\Msun}$. Thus, we do not need any gas mass contribution to explain our results. By taking the extreme 1$\sigma$ uncertainties on our mass estimates, we find an upper limit on the gas mass in the lens of $M_{\rm gas}(<\theta_{\rm Ein}) = 0.8 \times 10^{11} \unit{\Msun}$, consistent with recent estimates of the gas mass fraction in ETGs at $z\sim2$ of $5-10\%$ of the stellar mass \citep[e.g.][]{Magdis2021,Caliendo2021, Whitaker2021}. Given the predicted low gas mass, morphological quenching is unlikely. On the other hand, since the halo is ten times more massive than the critical mass of \SI{e12}{\Msun}, hot gas quenching \citep[e.g.][]{Cattaneo06, Gabor15} could be a potential mechanism that led to the formation of this ETG.

\subsection{Total mass density profile of the lens}
\label{sec:Properties/lens/slope}

\begin{figure}
    \centering
    \includegraphics[width=0.95\hsize]{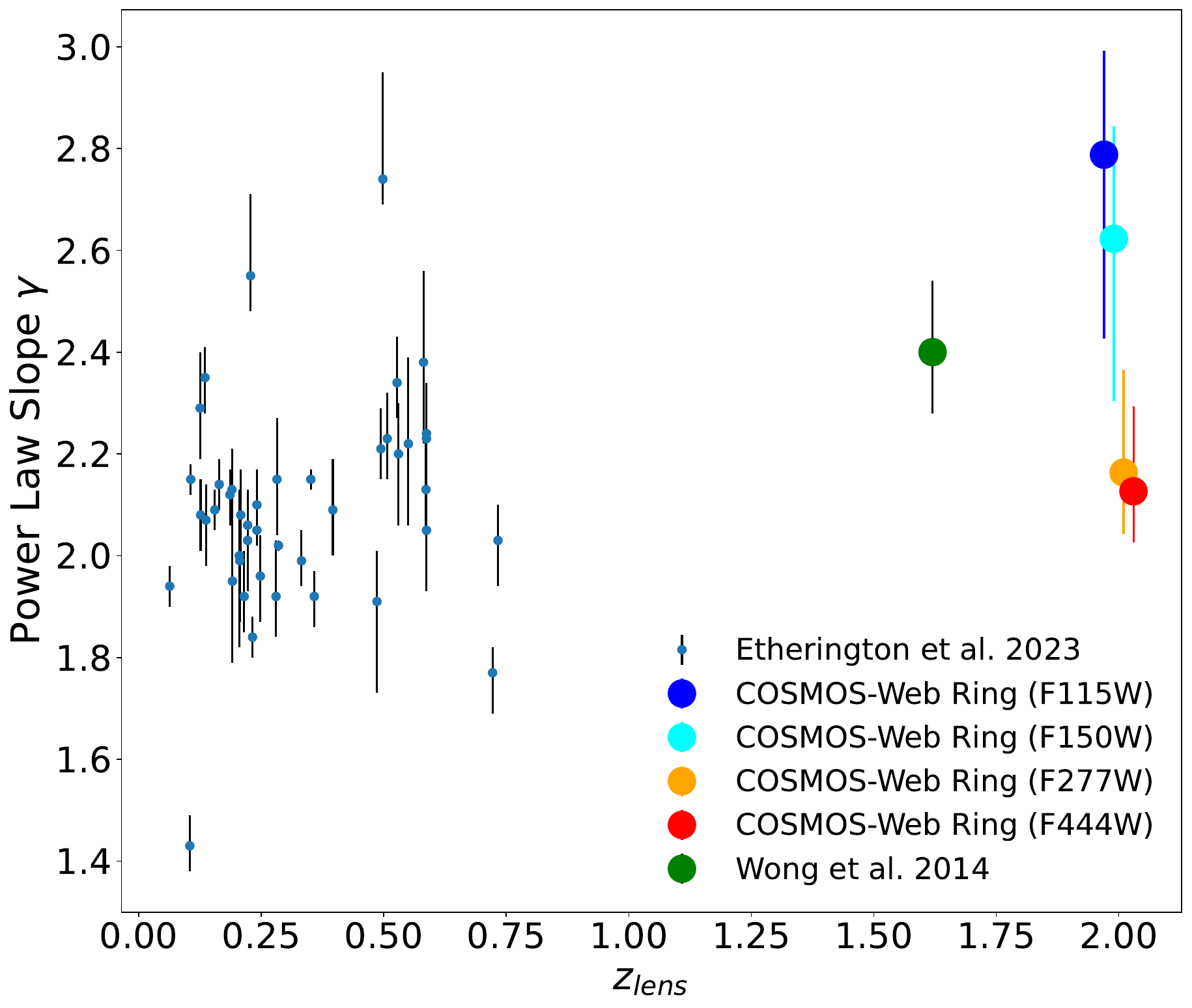}
    
    \caption{Inner slope $\gamma$ of the total mass profile of the lens as a function of redshift for 44 strong lenses from \citet{Etherington2023} (blue points) and the high redshift lens studied by \citet{Wong14} (green point). Independent values for the COSMOS-Web ring are shown for F115W (blue), F150W (cyan), F277W (orange), and F444W (red).}
    \label{fig:slopes}
\end{figure}

Measurements of the inner slope of the total mass density profile of ETGs inform us of how they evolve, for example the contribution of minor and major mergers and the role of processes like black hole feedback \citep[e.g.][]{Wang2019, Wang2020}. The slope of over 50 ETG strong lenses have been measured, with different correlations with redshift being claimed \citep[e.g.][]{Bolton2012, Sonnenfeld2013b, Etherington2023}. Existing statistically significant lens samples extend to redshifts of $z_{\rm lens} \sim 0.8$. The \Cring{} offers a first insight into the density slope of ETGs at $z \sim 2$.

Figure \ref{fig:slopes} compares the \Cring{}'s slope measurement to 44 values inferred in \citet{Etherington2023}, who used the same technique as this study. The measurement of \citet{Wong14} is also shown, which was the previous highest redshift slope measurement in a lens. All of our measured $\gamma$ values for the COSMOS ring are steeper than isothermal profiles ($\gamma = 2$), suggesting that higher redshift lenses may not have shallower density profiles than the local Universe average value around $\gamma = 2.06$ \citep{Koopmans2009}. However, with just a single lens and the large range of plausible $\gamma$ values measured, we cannot yet generalize our results to the broader population of ETGs.

Nevertheless, this study shows that lenses found via \JWST{} surveys like \COSMOSWeb{} will enable slope measurements extending to much higher $z_{\rm lens}$ values than previously. We also demonstrate that slope measurements is possible from the same JWST imaging data required to find the lens in the first place. Therefore, once the $\sim 50$ to $\sim 100$ lenses contained within the \COSMOSWeb{} data are found \citep{CWeb2022, Holloway2023}, this measurement will be possible on large lens samples.

The disparity between $\gamma$ measurements at bluer (F115W, F150W) and redder (F277W, F444W) wavelengths is noteworthy. The slope inferred via lens modeling depends on where the lensed source probes the mass distribution of the lens. In the bluer filters the source is brightest to the north and south of the lens, whereas in redder filters it is to the East and West. These different measurements might therefore indicate that the projected density of the lens varies azimuthally. This could be consistent with results from \citet{Nightingale2019} that showed that such effects could be possible because lenses may have distinct internal stellar structures. Ultimately, this would imply the underlying mass distribution of the lens is more complex than a power-law, as has been argued by other studies \citep{Schneider2014a, Etherington2023}. However, instrumental effects might also cause such effects given that different pixel scales are used between blue and red \JWST{} wavelengths and that the PSF shape and size significantly varies between those bands. Ideally, one should combine these lensing observations with stellar kinematics follow-ups to better probe the inner structure of the total mass distribution of the lens.

\subsection{Multiple galaxies in the source ?}
\label{sec:Properties/source/multiple galaxies}

When doing a source reconstruction, we find that the source has a highly irregular shape, as illustrated in Figs.\,\ref{fig:slfit_src} and \ref{fig:Lens model/PyAutoLens}. Given their differences in methodology, it is interesting to note that both \pyautolens{} and \slfit{} find similar results. The source seems to be made of at least two distinct components. The first one (Comp-1) is modeled as a compact galaxy oriented from East to West. It is mostly visible in F444W and it is the main contributor to \ClumpWest{} and \ClumpEast{} in the Einstein ring. Furthermore, its stellar mass derived on \slfit{} photometry is $M_\star \approx \SI{4e10}{\Msun}$. The second component (Comp-2) is more extended by roughly a factor of two with respect to Comp-1, oriented from North to South, and is offset to the East by about \SI{0.15}{\arcsec} (roughly \SI{1}{\kilo\pc}; see the bottom panel of Fig.\,\ref{fig:slfit_src} and right panel of Fig.\,\ref{fig:Lens model/PyAutoLens}). Comp-2 appears much less massive than Comp-1 with $M_\star \approx \SI{e9}{\Msun}$. When summed up, the stellar mass of Comp-1 and Comp-2 is consistent within the uncertainties with the stellar mass derived on the photometry of the whole ring from \slfit{} (see Table\,\ref{tab:Redshift/source}). As discussed in Sect.\,\ref{sec:Redshifts/source}, based on our fits with and without ground-based observations, a solution at $z_{\rm source} > 5$ is more likely. In particular, this solution appears very robust for Comp-2 compared to that of \citet{vandokkum2023massive} at $z_{\rm source} = 2.97$. However, we cannot easily discard the latter for Comp-1 given that its SED fits slightly better than the best-fit solution found at $z_{\rm{source}} = 5.48$ ($\chi^2 = 27$ instead of 34). A possibility could therefore be that the Einstein ring is actually the image produced by the superimposition along the line-of-sight of two galaxies at different redshifts. One way to determine whether that is the case or not is with follow-up observations using slit or integral field spectroscopy (e.g. X-shooter or NIRSpec IFU).

Assuming the source is located at a single redshift, then a second possibility is that the source is actually two galaxies in a merging process. The complex reconstructed morphology, the spatial offset between Comp-1 and Comp-2 by about \SI{1}{\kilo\pc}, and the differences in stellar mass, SFR, and color could be indications that there are two galaxies, potentially in interaction. If so, then the mass ratio between the two galaxies would be in the range of 1:15 - 1:75 given the uncertainties on the stellar masses of the two components which would correspond to a minor merger \citep[e.g.][]{Ventou2017, Ventou2019}. We note that this scenario would also be consistent with the fact that the component with the lowest stellar mass is the most extended one, for instance because of tidal stripping.

\subsection{A partially dust-obscured star-forming galaxy ?}
\label{sec:Properties/source/dust}

The source could also be a dust-obscured galaxy whose dust is inhomogeneously located throughout the galaxy. In particular, this could explain why the clumps \ClumpWest{} and \ClumpEast{} appear much redder than the rest of the ring as well as the highly irregular morphology, as suggested by both observations \citep[e.g.][]{Dye2015, Massardi2018} and simulations \citep[e.g.][]{Cochrane2019}. To check for the presence of dust, we have used FIR detection from Jin et al. (in prep.) as discussed in the last paragraph of Sect.\,\ref{sec:Observations/COSMOS-Web}. We do not get any detection in MIPS and \PACS{} bands (${\rm S/N} < 1$) and tentative detections with ${\rm S/N} \approx 1.5$ in the three \SPIRE{} bands when taking into account the confusion noise. 

However, we do measure a flux of \SI{4.9(1.1)}{\milli\jansky} in \SCUBA{} (${\rm S/N} \approx 4.3$). Its peak is not centered on the ring but offset to the South West, though the coarser resolution of \SCUBA{} (beam $\sim \SI{15}{\arcsec}$) make it difficult to determine its exact location. Because the lens is a passive elliptical galaxy, its FIR emission is expected to be low and not detectable in \SCUBA{}, as indicated by the stacked FIR SEDs of $z \sim 2$ quiescent samples \citep[e.g.][]{Magdis2021}. Besides, the mass analysis of the lens presented in Sect.\,\ref{sec:Properties/lens} is also consistent with little to no gas and therefore dust in the lens. In addition, \ClumpWest{} and \ClumpEast{} are the reddest objects detected in \NIRCAM{} within the \SCUBA{} beam, which is in favor of the ring being the origin of the dust emission. Still, the nearby companion at $z \approx 2$ could also contribute.

Within the Einstein ring, the two reddest components are \ClumpWest{} and \ClumpEast{} which could suggest that, if the background source is dusty, it is inhomogeneously distributed. This is supported by the fact that the best-fit SED model from \Lephare{} finds an attenuation of $E(B-V)=0.7$ for Comp-1 (i.e. for the clumps in the ring) but only $E(B-V)=0.1$ for Comp-2 which corresponds to the blue component of the ring. Another option is given by the fact that H$\alpha$ falls within the F444W band at $z_{\rm source} \sim 5.5$ in which case the difference in color between the clumps and the rest of the ring would be produced by star-formation. However, when comparing the SEDs of Comp-1 and Comp-2 we find that this explanation is unlikely since \Lephare{} was allowed to boost the emission line fluxes by up to a factor of two and never converged to such a solution. Therefore, the more likely scenario is that the background source is a dusty star-forming galaxy at $z > 5$ with an inhomogeneous distribution of dust.

Finally, we can also compare our stellar mass and SFR estimates to the Main Sequence (MS) of \citet{Khusanova2021}. Their MS was obtained from the \ALPINEALMA{} survey \citep{Bethermin2020, LeFevre2020, Faisst2020} by estimating with \ALMA{} the fraction of dust-obscured SFR in galaxies at $z \sim 4 - 5$ from their rest-frame FIR continuum. Taking their Fig.\,10, they find a ${\rm{SFR}} \approx \SI{30}{\Msun\per\year}$ for $M_\star = \SI{e10}{\Msun}$ and ${\rm{SFR}} \approx \SI{100}{\Msun\per\year}$ for $M_\star = \SI{5e10}{\Msun}$. Thus, the source falls within their MS. If the source is instead at $z_{\rm{source}} \approx 4.5$ and with $M_\star > \SI{e10}{\Msun}$, then it would rather lie just below their best-fit MS with a difference $\Delta {\rm{SFR}} \gtrsim \SI{100}{\Msun\per\year}$. We reach similar conclusions when comparing our results to the best-fit MS at \SI{1}{\giga\year} from \citet{Popesso2023}.

\section{Conclusions}
\label{sec:Conclusions}

We have presented in this paper an in-depth analysis of the \Cring{}. We serendipitously discovered it during the data reduction of the \COSMOSWeb{} survey in April 2023. Similarly, it has been independently discovered by another team and presented in \citet{vandokkum2023massive}. Our separate analysis leads to the following findings. The system comprises a central lens and a full Einstein ring with two red clumps noted CW and CE and mostly detected in the F444W band. A nearby companion is also located to the South West. It is found at $z \approx 2$ and is therefore likely associated with the lens. Thanks to the wealth of multi-wavelength observations in \COSMOSWeb{}, we combined our \JWST{} data with ground- and space-based observations from the visible to the FIR domain. Besides \JWST{}, the \Cring{} is also detected in \HST{}/F814W, \UVISTA{}, and \HSC{}-$i$ bands. However, these previous observations lacked sufficient resolution and S/N to identify the lens. Thus, the \Cring{} was effectively unnoticed by previous strong lens catalogues prior to the advent of \JWST{} observations in \COSMOSWeb{}.

By combining more than 25 bands from the u-band to the NIR and using robust model fitting techniques, we have extracted the photometry of both the lens and the source. This allowed us to derive the photometric redshifts and the physical properties of the lens and the source with three different SED fitting codes (\Cigale{}, \Lephare{}, and \EAZY{}). For the lens, we find consistent results that make it a red, compact (Sérsic index $n = 6.4$), massive ($M_{\star} \approx 1.37^{+0.14}_{-0.11} \times \SI{e11}{\Msun}$), and quiescent (${\rm{sSFR}} \lesssim \SI{e-13}{\Msun\per\year}$) ETG at $z_{\rm{lens}} = 2.02 \pm 0.02$. Given its size ($R_{\rm eff} = \SI{1.5}{\kilo\pc}$), it falls on the typical mass-size relation found for ETGs at the same redshift. For the source, our results also consistently show that it is a massive ($M_{\star} \approx 1.26^{+0.17}_{-0.16} \times \SI{e10}{\Msun}$) and star-forming galaxy (${\rm{SFR}} = 77.6^{+15.4}_{-11.0}~\unit{\Msun\per\year}$) at $z_{\rm{source}} \approx 5.48 \pm 0.06$. These values are consistent with those from \citet{vandokkum2023massive}, except for the redshift of the background source that they find at $z_{\rm{source}} \approx 2.93$ instead. Overall, we find no evidence for a solution at $z_{\rm{source}} < 3$, except when fitting the red component of the ring alone, in which case both $z_{\rm{source}} \approx 3$ and $z_{\rm{source}} \approx 5.5$ appear as valid solutions.

In addition, we have also carried out two different and complementary lens modelings on \JWST{} images. The first one is \slfit{} and uses a parametric approach to model the morphology of the source whereas the second one is \pyautolens{} and uses a pixel-grid reconstruction technique. This has allowed us to (i) constrain the mass budget of the lens by measuring its total mass within the Einstein radius, (ii) provide a first constraint of the inner slope of the total mass density profile of an ETG at $z \sim 2$, and (iii) reconstruct the complex morphology of the source. Both techniques provide fairly consistent results where the reconstructed morphology of the source is complex and highly waveband-dependent with at least two distinct components offset from each other. The best-fit mass profile slope $\gamma$ from \pyautolens{} systematically goes down from F115W to F444W with $\gamma = 2.78^{+0.20}_{-0.36}$ in F115W and $\gamma = 2.12^{+0.10}_{-0.17}$ in F444W. While a single lens is not sufficient to constrain the evolution of the slope of the total mass profiles of ETGs since $z \sim 2$, the \Cring{} is always found to have a value above the average $\gamma = 2.06$ found in the local Universe.

We analyzed the contribution of the baryonic and dark matter components to the total mass budget of the lens by comparing: \begin{enumerate*}[label=(\roman*)]
    \item the stellar mass measured from SED fitting, 
    \item the halo mass expected from SHMR relations, and 
    \item the total mass inferred from lensing. 
\end{enumerate*}
This mass budget is established within the Einstein radius.
Our results consistently show that the total mass budget can be fully accounted for by the measured stellar and dark matter masses. These conclusions are robust and hold for any set of stellar mass estimates and SHMR that we tested. We conclude that the ETG is likely hosted by a DM halo mass of $M_{\rm h} = 1.09^{+1.46}_{-0.57} \times \SI{e13}{\Msun}$. We also conclude that we do not need any gas mass contribution with an upper-limit of $M_{\rm gas}(<\theta_{\rm Ein}) = \SI{0.8e11}{\Msun}$. Therefore, in contrast to the findings in \citet{vandokkum2023massive} we report no missing mass and no need to change the IMF or the DM halo profile to interpret it. We find that the main reason behind our two opposite conclusions is that \citet{vandokkum2023massive} estimate a total mass a factor of two higher than ours because of their lower redshift for the source.

The morphology of the source reconstructed from the lens modeling is fairly consistent between the two methods we adopted. In both cases, at least two different components are necessary to account for the complex ring shape. The first one is relatively compact, accounts for the bulk of the stellar mass in the ring, and is the main contributor to \ClumpWest{} and \ClumpEast{}. The second component is about twice as extended as the first component, is offset to the West by roughly \SI{1}{\kilo\pc} and mostly emits in F115W and F150W. The source is likely a single galaxy at $z_{\rm{source}} \approx 5.5$ but, given the complex reconstructed morphology and that the solution at $z_{\rm{source}} = 2.93$ is also plausible for the red component, we cannot discard the possibility that the source is the superimposition along the line-of-sight of two galaxies at $z \approx 3$ and 5.5. If so, we estimate an upper limit on the probability to happen of 0.1\%. Furthermore, without spatially resolved spectroscopy (e.g. NIRSpec IFU or \ALMA{}), we also cannot discard the possibility that the source is actually a merger. If so, then the mass ratio of the two components would suggest that the source is a minor merger, with potentially one of the two galaxies being tidally stripped.

Finally, by cross-correlating the position of the \Cring{} with the Super-deblended catalogue of Jin et al. (in prep.), we find a nearby detection at ${\rm S/N} = 4.3$ in \SCUBA{} and tentative detections in \Herschel{}/\SPIRE{} bands. Given the position of the \SCUBA{} beam, the passive nature of the lens, and the large dust attenuation found by \Lephare{} in \ClumpWest{} and \ClumpEast{}, we estimate that the FIR detections likely originate from the clumps. This would make the background source a dusty star-forming galaxy with a potentially heterogeneous dust distribution.

This system offers exciting opportunities for studying star formation in a resolved galaxy during the first billion years of the Universe.

\begin{acknowledgements}
This research is also partly supported by the Centre National d'Etudes
Spatiales (CNES). 
We acknowledge the funding of the French Agence Nationale de la Recherche for the project iMAGE (grant ANR-22-CE31-0007). 
We warmly acknowledge the contributions of the entire COSMOS collaboration consisting of more than 100 scientists. The HST-COSMOS program was supported through NASA grant HST-GO-09822. More information on the COSMOS survey is available at \url{https://cosmos.astro.caltech.edu}.
This work was made possible by utilizing the CANDIDE cluster at the Institut d’Astrophysique de Paris, which was funded through grants from the PNCG, CNES, DIM-ACAV, and the Cosmic Dawn Center and maintained by S. Rouberol. 
JWN is supported by the UK Space Agency, through grant ST/N001494/1. JR was supported by JPL, which is operated by Caltech under a contract for NASA. 
GEM acknowledges the Villum Fonden research grant 13160 “Gas to stars, stars to dust: tracing star formation across cosmic time,” grant 37440, “The Hidden Cosmos,” and the Cosmic Dawn Center of Excellence funded by the Danish National Research Foundation under the grant No. 140. 
SJ is supported by the European Union’s Horizon Europe research and innovation program under the Marie Sk\l{}odowska-Curie grant agreement No. 101060888.
JN, GM and RM are supported by STFC via grant ST/X001075/1, and the UK Space Agency via grant ST/X001997/1. AA is supported by the European Research Council via grant DMIDAS (GA 786910). This work used both the Cambridge Service for Data Driven Discovery (CSD3) and the DiRAC Data-Centric system, which are operated by the University of Cambridge and Durham University on behalf of the STFC DiRAC HPC Facility (www.dirac.ac.uk). These were funded by BIS capital grant ST/K00042X/1, STFC capital grants ST/P002307/1, ST/R002452/1, ST/H008519/1, ST/K00087X/1, STFC Operations grants ST/K003267/1, ST/K003267/1, and Durham University. DiRAC is part of the UK National E-Infrastructure.
\end{acknowledgements}

\bibliographystyle{aa}
\bibliography{biblio.bib}

\appendix

\onecolumn
\section{Additional figures}

\begin{figure*}[hbtp]
    \centering

    \includegraphics[width=\hsize]{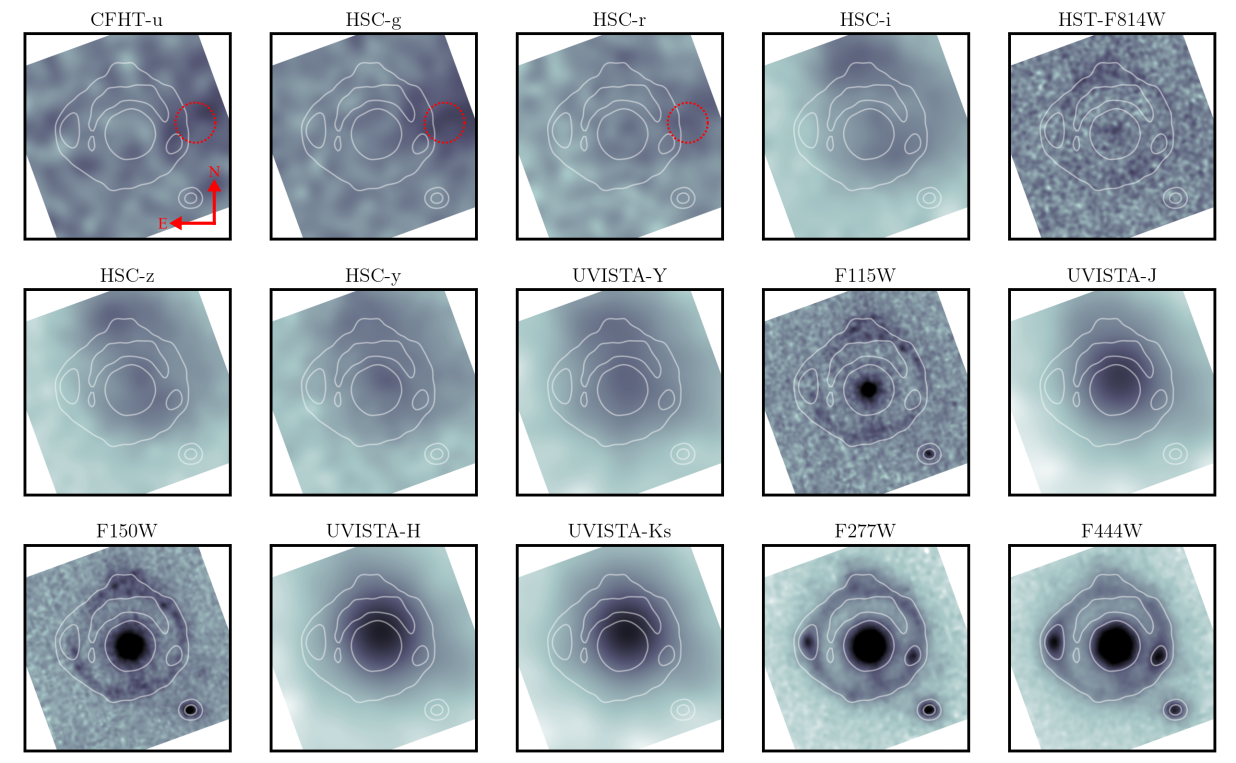}
    
    \caption{Cutouts of  $\times$ 3 \unit{\arcsec} of the Einstein ring in all bands. Images are sorted from the bluest to the reddest band going left to right and top to bottom. Each image is scaled using a square root function and \texttt{ZScale} intervals to enhance the contrast between the bright lens and the fainter surrounding ring. Contours from the detection image of \sextractor{} are shown in all images as white lines. The circle with red dotted lines show the location of the nearby UV-bright contaminant discussed in Sect.\,\ref{sec:Observations/discovery} and in Fig.\,\ref{fig:Redshifts/SEDs}. North is up and East is left.
    }
    \label{fig:Appendix/cutouts}
\end{figure*}

\end{document}